\RequirePackage[2020-02-02]{latexrelease} 
\documentclass[twocolumn]{aastex631}
\submitjournal{ApJ}

\shorttitle{AS2COSPEC : survey description and first results}
\shortauthors{Chen et al.}

\usepackage{amsmath,amssymb}
\usepackage{chngcntr}

\begin{document}

\title{An ALMA Spectroscopic Survey of the Brightest Submillimeter Galaxies in the SCUBA-2–COSMOS field (AS2COSPEC): Survey Description and First Results 
}

\correspondingauthor{Chian-Chou Chen (TC)}
\email{ccchen@asiaa.sinica.edu.tw}

\author[0000-0002-3805-0789]{Chian-Chou Chen}
\affiliation{Academia Sinica Institute of Astronomy and Astrophysics (ASIAA), No. 1, Sec. 4, Roosevelt Rd., Taipei 10617, Taiwan}

\author{Cheng-Lin Liao}
\affiliation{Graduate Institute of Astrophysics, National Taiwan University, Taipei 10617, Taiwan}
\affiliation{Academia Sinica Institute of Astronomy and Astrophysics (ASIAA), No. 1, Sec. 4, Roosevelt Rd., Taipei 10617, Taiwan}

\author{Ian Smail}
\affiliation{Centre for Extragalactic Astronomy, Department of Physics, Durham University, South Road, Durham DH1 3LE, UK}

\author{A. M. Swinbank}
\affiliation{Centre for Extragalactic Astronomy, Department of Physics, Durham University, South Road, Durham DH1 3LE, UK}

\author{Y. Ao}
\affiliation{Purple Mountain Observatory and Key Laboratory for Radio Astronomy, Chinese Academy of Sciences, Nanjing, China}

\author{A. J. Bunker}
\affiliation{Department of Physics, University of Oxford, Keble Road, Oxford OX13RH, UK}

\author{S. C. Chapman}
\affiliation{Department of Physics and Astronomy, University of British Columbia, 6225 Agricultural Road, Vancouver V6T 1Z1, Canada}
\affiliation{National Research Council, Herzberg Astronomy and Astrophysics, 5071 West Saanich Road, Victoria V9E 2E7, Canada}
\affiliation{Eureka Scientific, Inc. 2452 Delmer Street Suite 100, Oakland, CA 94602-3017, USA}

\author{B. Hatsukade}
\affiliation{Institute of Astronomy, Graduate School of Science, The University of Tokyo, 2-21-1 Osawa, Mitaka, Tokyo 181-0015, Japan}

\author{R. J. Ivison}
\affiliation{European Southern Observatory, Karl Schwarzschild Strasse 2, D-85748 Garching, Germany}

\author[0000-0002-2419-3068]{Minju M. Lee}
\affiliation{Cosmic Dawn Center (DAWN)} 
\affiliation{DTU-Space, Technical University of Denmark, Elektrovej 327, DK2800 Kgs. Lyngby, Denmark}
\affiliation{Max-Planck-Institut f\"{u}r Extraterrestrische Physik (MPE), Giessenbachstr. 1, D-85748 Garching, Germnay}

\author{Stephen Serjeant}
\affiliation{School of Physical Sciences, The Open University, Milton Keynes, MK7 6AA, UK}

\author{Hideki Umehata}
\affiliation{Institute for Cosmic Ray Research, The University of Tokyo, 5-1-5 Kashiwanoha, Kashiwa, Chiba 277-8582, Japan}
\affiliation{Institute of Astronomy, The University of Tokyo, 2-21-1 Osawa, Mitaka, Tokyo 181-0015, Japan}

\author{Wei-Hao Wang}
\affiliation{Academia Sinica Institute of Astronomy and Astrophysics (ASIAA), No. 1, Sec. 4, Roosevelt Rd., Taipei 10617, Taiwan}

\author{Y. Zhao}
\affiliation{Yunnan Observatories, Chinese Academy of Sciences, Guandu District,Kunming 650011, People’s Republic of China}




\begin{abstract}
We introduce an ALMA band 3 spectroscopic survey, targeting the brightest submillimeter galaxies (SMGs) in the COSMOS field. Here we present the first results based on the 18 primary SMGs that have 870\,$\mu$m flux densities of $S_{870}=12.4-19.3$\,mJy and are drawn from a parent sample of 260 ALMA-detected SMGs from the AS2COSMOS survey. We detect emission lines in 17 and determine their redshifts to be in the range of $z=2-5$ with a median of ${3.3\pm0.3}$. We confirm that SMGs with brighter $S_{870}$ are located at higher redshifts. The data additionally cover five fainter companion SMGs, and we obtain line detection in one. Together with previous studies, our results indicate that for SMGs that satisfy our selection, their brightest companion SMGs are physically associated with their corresponding primary SMGs in $\ge40$\% of the time, suggesting that mergers play a role in the triggering of star formation. By modeling the foreground gravitational fields, $<10$\% of the primary SMGs can be strongly lensed with a magnification $\mu>2$. We determine that about 90\% of the primary SMGs have lines that are better described by double Gaussian profiles, and the median separation of the two Gaussian peaks is 430$\pm$40\,km\,s$^{-1}$. This allows estimates of an average baryon mass, which together with the line dispersion measurements puts our primary SMGs on the similar mass-$\sigma$ correlation found on local early-type galaxies. Finally, the number density of our $z>4$ primary SMGs is found to be $1^{+0.9}_{-0.6}\times10^6$\,cMpc$^{-3}$, suggesting that they can be the progenitors of $z\sim3-4$ massive quiescent galaxies.

\end{abstract}



\section{Introduction}\label{sec:intro}
Galaxies that are bright in far-infrared and submillimeter host sites of both vigorous star formation and active black hole growth \citep{Sanders:1988aa}. Predominantly at $z=1-4$ (e.g., \citealt{Chapman:2005p5778,Ivison:2007fj, Wardlow:2011qy, Casey:2014aa, Hodge:2020aa}), recent studies focusing on submillimeter galaxies that are bright at 850\,$\mu$m ($S_{\rm 850}\gtrsim1$\,mJy), or SMGs, found that they are some of the most massive systems during these epochs, sitting at the massive end of the stellar mass functions \citep{Dye:2008ys,Hainline:2011aa,Michaowski:2012fr,Koprowski:2016aa,Dudzeviciute:2020aa}, molecular gas mass functions \citep{Bothwell:2013lp,Birkin:2021aa}, dust mass functions \citep{da-Cunha:2015aa,da-Cunha:2021aa}, and also the halo mass functions \citep{Hickox:2012kk,Chen:2016ab,Wilkinson:2017aa,An:2019aa,Stach:2021aa}. The large cold molecular gas reservoir ($\sim10^{11}$\,M$_\odot$) estimated from their molecular line measurements is believed to provide the gas supply over $\sim$100\,Myr timescales for their extensive star-formation rates (SFRs) of $\sim$100-1000\,M$_\odot$  yr$^{-1}$, similar to the levels of star formation seen in nearby (hyper-)ultra-luminous infrared galaxies (HyLIRGs/ULIRGs; \citealt{Sanders:1996p6419}). Supported by observational evidence including large-scale clustering, the current hypothesis is that SMGs are intimately linked to luminous quasars at similar redshifts, and likely followed by phases of compact quiescent galaxies and the local massive ellipticals \citep{Lilly:1999lr,Swinbank:2006aa,Alexander:2012aa,Toft:2014aa,Dudzeviciute:2020aa}. The fact that the massive ellipticals dominate the stellar mass budget among the local massive galaxies ($\gtrsim M_\ast$; \citealt{Ilbert:2013aa}) suggests that the phases of SMGs could be responsible for the formation of most of their stellar masses \citep{Thomas:2010aa}.

Their essential role in formation and evolution of massive galaxies highlights how incredibly little we understand SMGs even after over two decades since their first discovery \citep{Smail:1997p6820,Barger:1998p13566,Hughes:1998p9666}. Technical difficulties compounded with their extreme infrared luminosity have kept many issues unsettled regarding the basic physical properties of SMGs, both in theory and observations. The one issue, and arguably the only one, that has recently been considered more or less clear, is the SMG number densities, or the number counts. Observational results have now converged to within statistical uncertainties over a wide flux range after taking into account field-to-field variations \citep{Karim:2013fk,Simpson:2015ab,Oteo:2016aa,Geach:2017aa,Hill:2018aa,Stach:2018aa,Simpson:2020aa}. However this comes only after years of efforts in examining the impact of multiplicity\footnote{Submillimeter sources uncovered by single-dish observations breaking into multiple constituent SMGs in follow-up interferometric observations.} on counts deduced from single-dish measurements, which typically served as the basis of constructing the number counts \citep{Wang:2011p9293,Barger:2012lr,Chen:2013gq,Hodge:2013lr,Cowie:2018aa}. Models have also made significant progress, where measured number counts can now be reproduced, in many cases without having to invoke top-heavy stellar initial mass functions \citep{Lacey:2016aa,Safarzadeh:2017aa,Bethermin:2017aa,Casey:2018aa,Lagos:2020aa,Lovell:2021aa}. {Although note that top-heavy stellar initial mass functions have been suggested in recent observational studies \citep{Zhang:2018aa,Dye:2022aa}.} However, a variety of different techniques used to construct models has led to vastly different predictions on some of the most fundamental properties.


\begin{figure*}[ht!]
	\begin{center}
		\leavevmode
		\includegraphics[scale=0.9]{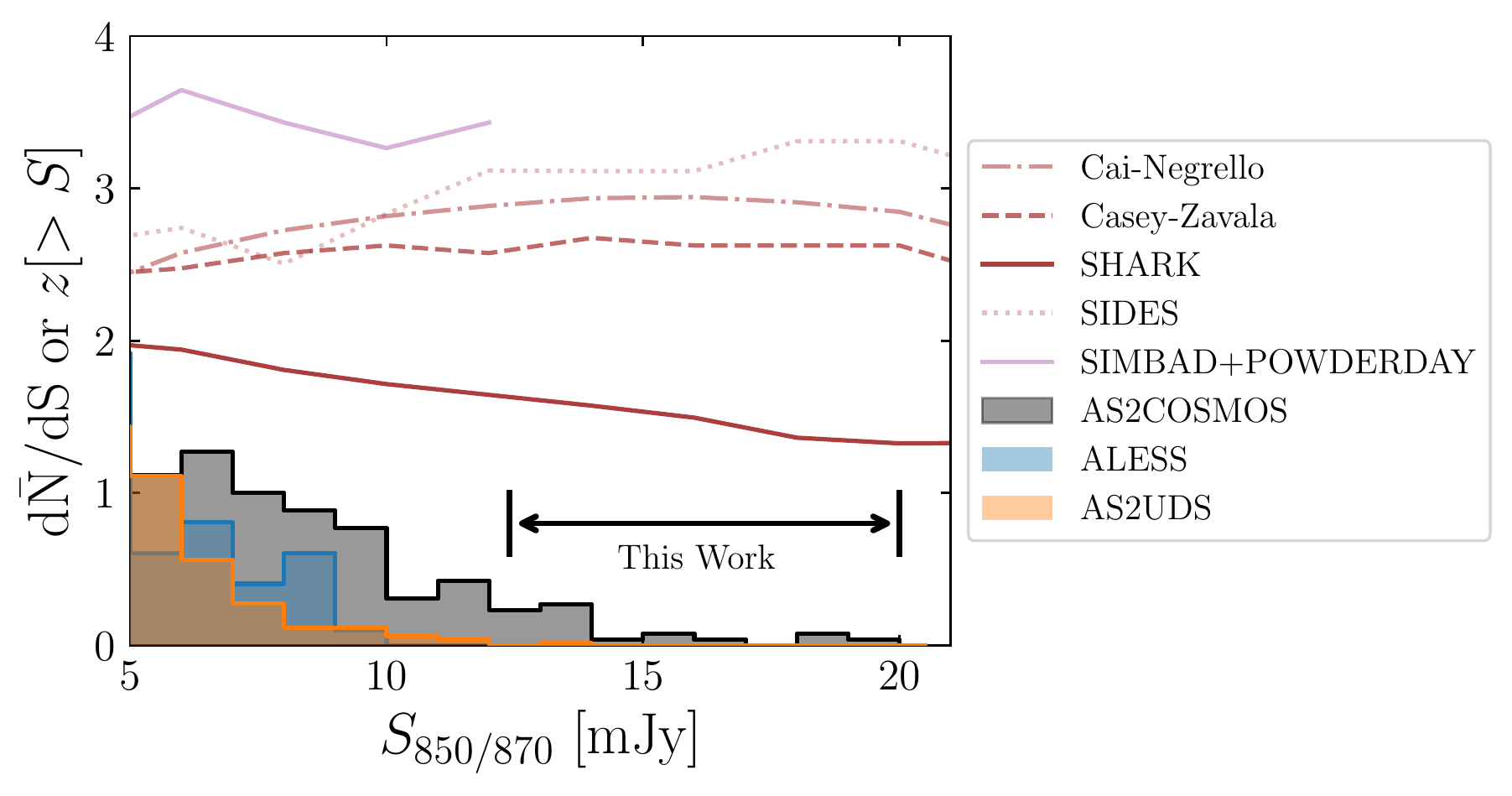}
		\caption{Predictions of median redshifts for SMGs with 850\,$\mu$m flux density ($S_{850}$) greater than certain flux cuts, based on empirical models of Cai-Negrello \citep{Cai:2013aa,Negrello:2017aa}, Casey-Zavala \citep{Casey:2018aa,Zavala:2021aa}, SIDES \citep{Bethermin:2017aa}, the semi-analytical model SHARK \citep{Lagos:2020aa}, and the hydro-dynamical simulation SIMBA \citep{Dave:2019aa} coupled with dust radiative transfer modeling POWDERDAY \citep{Narayanan:2021aa,Lovell:2021aa}. In the bottom of the figure we show the histograms normalized of three SMG samples that are constructed via a similar approach, meaning ALMA band 7 continuum follow-up observations on flux limited samples of single-dish detected submillimeter sources. The histograms are arbitrarily normalized to show contrasts of different samples. AS2COSMOS \citep{Simpson:2020aa} has significantly more brighter SMGs than AS2UDS \citep{Stach:2018aa} and ALESS \citep{Hodge:2013lr}. The flux range presented in this work is indicated, which is ideal for redshift measurements thanks to its brightness and test model predictions.  
		}
		\label{fig:fig1}
	\end{center}
\end{figure*}

One outstanding example is redshift; While most models can reproduce the average redshift distribution (median of $z\sim2.5$; \citealt{Chapman:2005p5778}) of typical SMGs with 850\,$\mu$m fluxes of $S_{850}\sim2-9$\,mJy, they predict opposite trends and correlations toward the brighter end; A school of models predict a median redshift of $z<2$ at $S_{850}\gtrsim9$\,mJy \citep{Lacey:2016aa,Lagos:2020aa} and another a median $z>3$ (\autoref{fig:fig1}; \citealt{Bethermin:2017aa,Casey:2018aa,Lovell:2021aa}). In addition, some models predict more or less a linear correlation between median redshift and $S_{850}$ but others have no such correlation. The differences in these predictions are likely due to different treatments of physical processes such as the triggering of star formation and dust formation, or the adoption of phenomenological recipes driven by the observations themselves. Precise redshift measurements of bright SMG samples that are sufficiently large and complete would be powerful in testing these predictions, and therefore represent the next frontier in understanding the formation of massive galaxies across cosmic time. 

In the pre-ALMA era, redshift measurements are focused on SMGs that in many cases have radio or mid-infrared detection, which was the technique used to identify their optical and near-infrared (OIR) counterparts after being discovered from single-dish submillimeter surveys \citep{Ivison:1998p10286,Barger:1999p6801,Smail:2000p6377,Chapman:2003ti,Chapman:2005p5778,Ivison:2007fj,Wardlow:2011qy,Umehata:2014aa}. Their redshifts were mostly estimated via simple flux ratios and in some cases OIR spectroscopic observations. They typically found a wide range of {median redshifts between $z=2-3$, with uncertainties on the order of 10-20\%}. In the ALMA era, where precise locations can be efficiently measured and counterpart identification is less biased, studies using well-defined samples have found more stable results of an overall median of $z\sim2.5-2.6\pm0.1$, with a moderate increase in median as a function of flux density \citep{Simpson:2014aa,Cowie:2017aa,Stach:2019aa}. However majority of these redshift are {estimated} via fittings of spectral energy distributions (SEDs), thus any possible systematic offsets due to model assumptions need to be understood with spectroscopic redshift measurements.

Spectroscopic redshift measurements of SMGs are known to be difficult; With about two hours exposures from 8-10\,m class OIR telescopes, the line detection rates were found to be about $\sim$30\% \citep{Casey:2017aa,Cowie:2017aa,Danielson:2017aa}. (Sub-)Millimeter spectral-scan observations targeting bright CO and [CI] lines appear more feasible for SMGs that are dusty but ISM rich. Indeed, with moderate time investment, this strategy has been successfully demonstrated on samples of very bright and typically strongly lensed sources discovered by {\it Herschel} or the South Pole Telescope \citep{Vieira:2013vn,Strandet:2016aa,Neri:2020aa,Bakx:2020aa,Reuter:2020aa,Urquhart:2022aa}. However, a generally lack of these very bright ($S_{850}\gtrsim25$\,mJy) SMGs in models motivates similar observations to be conducted on fainter samples which are less influenced by gravitational lensing  \citep{Birkin:2021aa,Jones:2021aa}. 

\begin{figure*}[ht!]
	\begin{center}
		\leavevmode
		\includegraphics[scale=0.6]{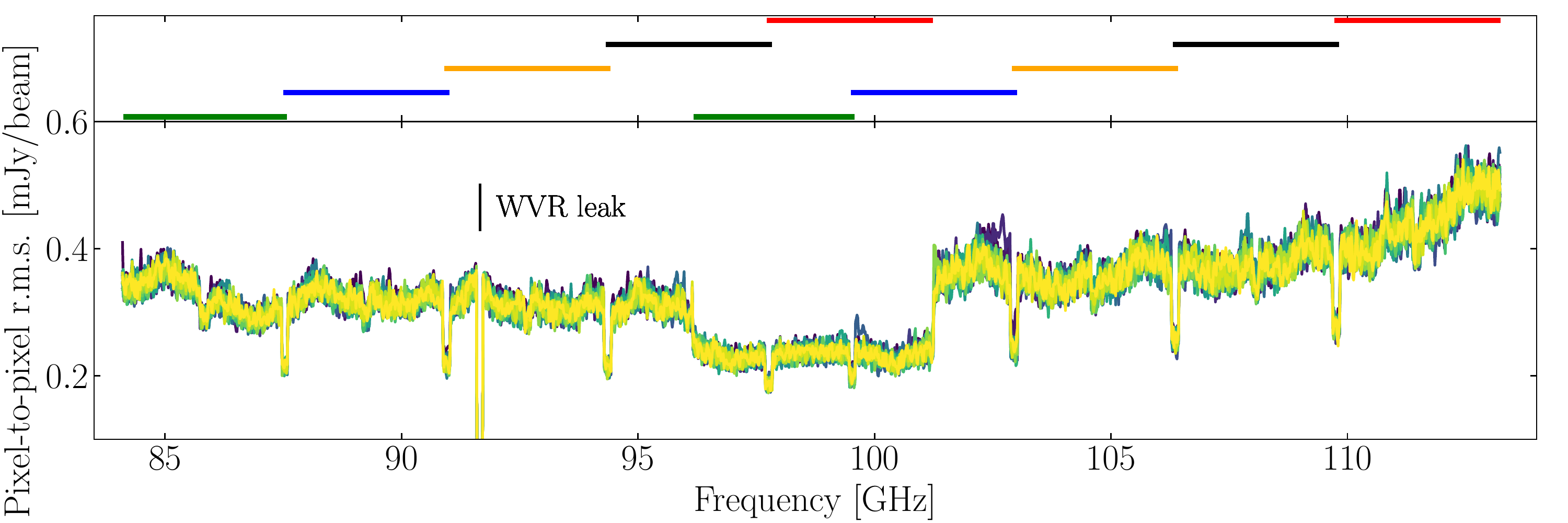}
		\caption{{\it Top}: The spectral coverage of the five sets of tuning adopted for our ALMA observations, where each set is plotted the same color. {\it Bottom:} The spectral sensitivity curves of our ALMA data expressed in pixel-to-pixel r.m.s., which are made based on data cubes smoothed to a common uniform beam size and shape. The curves of all 18 sources are plotted with each being assigned to a random brightness of green, and this is to show the small scatter (a few percent) of data quality among each of our sample source thanks to the observing strategy. The overall r.m.s. median is 0.33\,mJy\,beam$^{-1}$ per 15.6\,MHz ($\sim40-55$\,km\,s$^{-1}$), with notable dips due to overlaps of spectral coverage between different tuning, except for the marked position due to the WVR leak, which was flagged during the QA2 process. 
		}
		\label{fig:fig2}
	\end{center}
\end{figure*}

Recently, \citet{Simpson:2020aa} published a AS2COSMOS catalog of 260 SMGs that is based on sub-arcsecond ALMA band 7 continuum follow-up observations of the brightest 182 single-dish detected submillimeter sources ($S_{850}>6.2$\,mJy) drawn from the parent 1.6\,degree$^2$ of the SCUBA-2 850\,$\mu$m survey in the COSMOS field (S2COSMOS; \citealt{Simpson:2019aa}). Crucially, considering the depths of both the initial SCUBA-2 survey and the ALMA follow-up observations, the catalog provides $\sim$20 and $\sim$60 SMGs that are 100\% and $>$90\% complete, at flux cuts of $S_{850}\sim12$\,mJy and $S_{850}\sim9$\,mJy, respectively (\autoref{fig:fig1}; Simpson et al. 2020). Under similar flux cuts the bright sub-samples are a factor of $>3$ and $>50$ larger than those of AS2UDS \citep{Stach:2019aa} and ALESS \citep{Hodge:2013lr}, the two previous largest uniform ALMA follow-up SMG surveys, representing a major step forward in constructing a sizable and highly complete bright SMG sample. Together with the fact that the model predictions differ the most in this regime, the brightest SMGs (effectively high-redshift HyLIRGs) from AS2COSMOS provides an ideal testbed for constraining models (\autoref{fig:fig1}). 
 
Here we present the first results of an ALMA spectroscopic survey of the brightest AS2COSMOS SMGs, or AS2COSPEC. We present sample selection and the ALMA data in \autoref{sec:obs}. In \autoref{sec:results} we present spectral extraction, as well as analyses on detailed modeling of the line profiles and redshift determinations. The implications and comparisons of our results are discussed in \autoref{sec:discussion} and a summary is given in \autoref{sec:sum}. Note throughout this paper $S_{850}$ and $S_{870}$ are used to represent the flux measurements at the representative wavelengths of SCUBA-2 and ALMA band 7 observations, respectively. In reality these measurements can be used interchangeably given the current precision of flux measurements. We assume the {\it Planck} cosmology: H$_0 =$\,67.7\,km\,s$^{-1}$ Mpc$^{-1}$, $\Omega_M = $\,0.31, and $\Omega_\Lambda =$\,0.69 \citep{PlanckCollaboration:2020aa}. 

\section{Observations and data} \label{sec:obs}

\subsection{Sample} \label{sec:alma2018}
Our sample is drawn from AS2COSMOS, a parent sample of 260 SMGs identified by ALMA band~7 continuum follow-up observations of a highly complete (99.5\%), flux-limited ($S_{\rm 850}>6.2$\,mJy) sample of the brightest submillimeter sources uncovered by the single-dish SCUBA-2 survey at 850\,$\mu$m in the COSMOS field over an area of 1.6 deg$^2$ (Simpson et al. 2019; 2020).  The ALMA band~3 observations presented in this paper targeted the brightest 18 AS2COSMOS SMGs that are at the flux range of $S_{870}=12.4-19.3$\,mJy based on the ALMA measurements \citep{Simpson:2020aa}. The main goals are to measure the redshift distributions, gravitational lensing magnifications, as well as to study the ISM properties via molecular lines such as CO and [CI]. To compare to the previous large samples of ALMA-identified SMGs from follow-up observations of single-dish detected submillimeter sources, in \autoref{fig:fig1} we show that there are only two SMGs that are similarly bright as our sample in the AS2UDS sample \citep{Stach:2018aa} and there is none in the ALESS sample \citep{Hodge:2013lr}. Note that {about one third} of the sample was observed in millimeter by NOEMA and other ALMA datasets, and their measured redshifts were reported by \citet{Jimenez-Andrade:2020aa}, \citet{Simpson:2020aa}, and \citet{Birkin:2021aa}, which are consistent with our results.

\subsection{ALMA data} \label{sec:alma2019}

The data were taken in November and December of 2019, under the project number 2019.1.01600.S. The observations were split into eleven execution blocks, each of which amounts about one and half hours worth of data for both science and calibrations. The default spectral scan setup was adopted in the {ALMA Observing Tool (OT)}, meaning that each execution block contains data on all 18 SMGs with five spectral tunings covering continuously from 84.1 to 113.3 GHz (\autoref{fig:fig2}), where each tuning with four spectral windows was carried out for exactly 26s. That is, across the whole frequency range observed, each spectral channel received totally about 4.8 minutes (26s times 11 executions) of science data on each SMG, except in the range of 97.7--101.3\,GHz, as well as overlapping edge channels, where the on-target exposure time roughly doubles. Each spectral window is tuned to a standard {Time Division Mode (TDM)} with 128 channels and a frequency width of 15.625 MHz, corresponding to a velocity width of about 40--55\,km\,s$^{-1}$ depending on the observed frequency. At least 43, up to 49, antennas were used for each execution block, with a baseline length ranging from 15 to 313 meters hence a nominal C43-2 configuration. Since the target SMGs are all located in the COSMOS field, the phase calibrator was always J1008+0029, and J1037-2934 was always used for the amplitude and bandpass calibrations. The weather conditions during these observations were standard band 3 weather, with a {precipitable water vapor (PWV)} ranging from 1.7 to 5.8 mm.

We adopt the calibrations performed {in the second level of the quality assurance (QA2)}. Flagging and calibrations were done using {\sc casa} pipeline version 5.6.1-8, which is also the version used for imaging. Additional manual flags were put in by the QA2 team to remove antennas or scans with poor performance, in particular a line at 91.66 GHz leaking from the {local oscillator (LO)} water vapor radiometer, which is marked in \autoref{fig:fig2}. We reviewed the weblog record and confirmed the quality of the calibration results. 

To image the visibilities, the calibrated visibilities were first continuum subtracted. To do that, the channels that contain line emissions need to be excluded for a proper first-order continuum fitting. The line channels to be masked were determined from the spectra extracted from the delivered pipeline-produced imaging cubes through typical single Gaussian fitting, spanning across +/- 3 sigma centered at the Gaussian peaks based on the fits. The spectra were extracted using circular apertures centered at the 870 micron continuum positions (the phase centers), and their radius were determined by a curve-of-growth method. 

\begin{figure}[ht!]
	\begin{center}
		\leavevmode
		\includegraphics[scale=0.25]{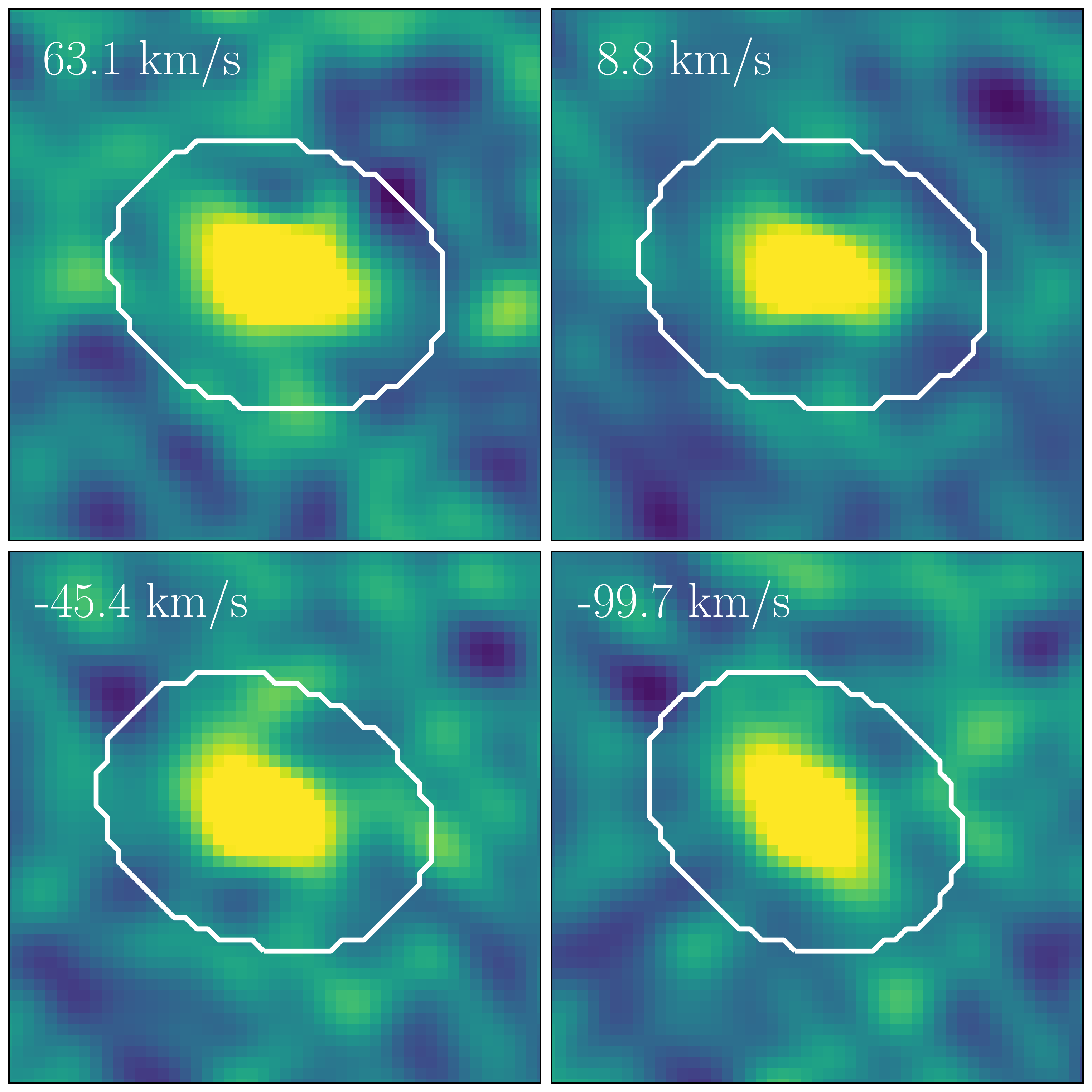}
		\caption{An example of channel maps of AS2COS0023.1, each 20$''$ on a side, where the velocities are referenced to the systemic redshift. The masked regions for {\sc tclean} are enclosed within the white curves. This demonstrates the flexibility of auto-masking in {\sc tclean}.
		}
		\label{fig:fig3}
	\end{center}
\end{figure}

\begin{figure*}[ht!]
	\begin{center}
		\leavevmode
		\includegraphics[scale=0.7]{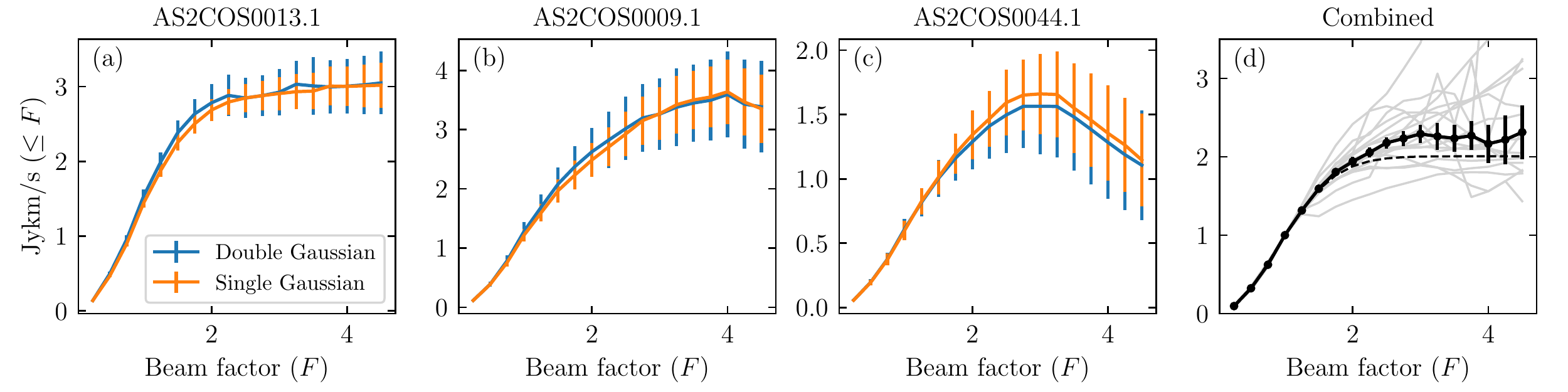}
		\caption{Example results of the curve-of-grown analyses, showing the curves converging (a), monotonically increasing (b) , and down turning (c), as a function of the size of the synthesized beam. Results from the single and double Gaussian models are both shown, as indicated in the legend. In panel (d) we show that by normalizing all curves to a beam factor of one (grey), the median curve (solid) with the bootstrapped uncertainties appears stable and converging, which is consistent with the shape of the synthesized beam (dashed) and showing that the non-converging behavior of individual targets could be due to larger-scale noise structures. We therefore in principle adopt the synthesized beam as the aperture to extract spectra and use this median curve to correct the measured flux densities to total. Note that this median curve is stable to within a few percent against the choice of aperture size to which all the curves are normalized.
		}
		\label{fig:fig4}
	\end{center}
\end{figure*}

We then used {\sc Tclean} to inverse Fourier transform and clean the continuum-subtracted visibilities. The visibilities were transformed to image cubes of 270$\times$270 pixels in {\it x} (RA) and {\it y} (DEC) axis with a pixel size of 0$\farcs$42 ($\sim$6-8 pixels per synthesized beam), and 1870 channels in the {\it z} (frequency) axis. We adopt natural baseline weighting across all spectral windows in order to maximize the signal-to-noise ratio (SNR) for line detection. To increase the efficiency of clean, we adopted the auto-masking approach \citep{Kepley:2020aa} in {\sc Tclean}, which allows {\sc casa} to generate masks for each channel in an iterative process based on a few parameters. To do that we set the {\it usemask} parameter to 'auto-multithresh'. For each channel, the first round of mask creation only includes pixels with SNR of four and above ({\it noisethreshold} = 4), and then the mask is allowed to expand in order to include lower signal-to-noise regions ({\it lownoisethreshold} = 2.5). The masked regions determined by auto-masking were then cleaned down to 2 sigma level (nsigma = 2 in {\sc Tclean}). Examples of the cleaned channels and their associated masks are shown in \autoref{fig:fig3}. Finally, given the same baseline weighting across the whole frequency range, the spatial resolution of each frequency channel differs slightly, up to $\sim$35\% end-to-end. To allow more straightforward analyses and understanding of the data we applied smoothing on the reduced cubes by using the CASA routine {\sc imsmooth}, setting the kernel to the common resolution, which is normally the largest beam size of the pre-smoothed cube. After this step, the spatial resolution of all cubes is uniform, with a synthesized beam FWHM of 4$\farcs$3$\times$3$\farcs$5 and a small P.A. range of 69-72 degrees. 
The final spectral sensitivity achieved is uniform across all sources to within a few percents thanks to the observing strategy, however it differs among channels mainly due to the tuning coverage. We show the pixel-to-pixel r.m.s. in \autoref{fig:fig2}, and the overall median is 0.33 mJy\,beam$^{-1}$ per 15.6 MHz ($\sim$40-55\,km\,s$^{-1}$). 
The continuum of each source was also imaged with natural weighting, resulting into images with a synthesized beam FWHM of 3$\farcs$6$\times$3$\farcs$0 and a small P.A. range of 77-80 degrees. The sensitivity achieved for continuum is also uniform across different sources, 12-13 $\mu$Jy\,beam$^{-1}$ at the representative frequency 98.7 GHz.

\section{Analyses and Results} \label{sec:results}
\subsection{Line detection and spectra extraction}\label{sec:spec}
To systematically search for line emission and quantify their significance, we ran the publicly available code {\sc LineSeeker}, which was first written and used to search for line emission for the ALMA frontier field survey \citep{Gonzalez-Lopez:2017aa}, and was later applied to the data taken for the ALMA Spectroscopic Survey in the Hubble Ultra Deep Field (ASPECS; \citealt{Walter:2016aa, Gonzalez-Lopez:2019aa}). {\sc LineSeeker} utilizes a matched filtering technique that combines spectral channels based on Gaussian kernels with a range of widths. The widths are chosen to match lines detected in real observations. The line candidates are determined to be significantly detected if their signal-to-noise ratios are larger than the most significantly detected negative signals in the cube{, and we find a significance threshold of SNR$>$10 satisfies the aforementioned requirement in all cases.} In this paper we focus on the spectra of the known SMGs in the AS2COSMOS catalog (Simpson et al. 2020), and the results from the search of serendipitous detection will be presented in a future work. For the 18 primary SMGs, 17 have yielded significant line detection based on {\sc LineSeeker}, and AS2COS0037.1 is the only source that has none. 

\begin{figure*}[ht!]
	\begin{center}
		\leavevmode
		\includegraphics[scale=0.95]{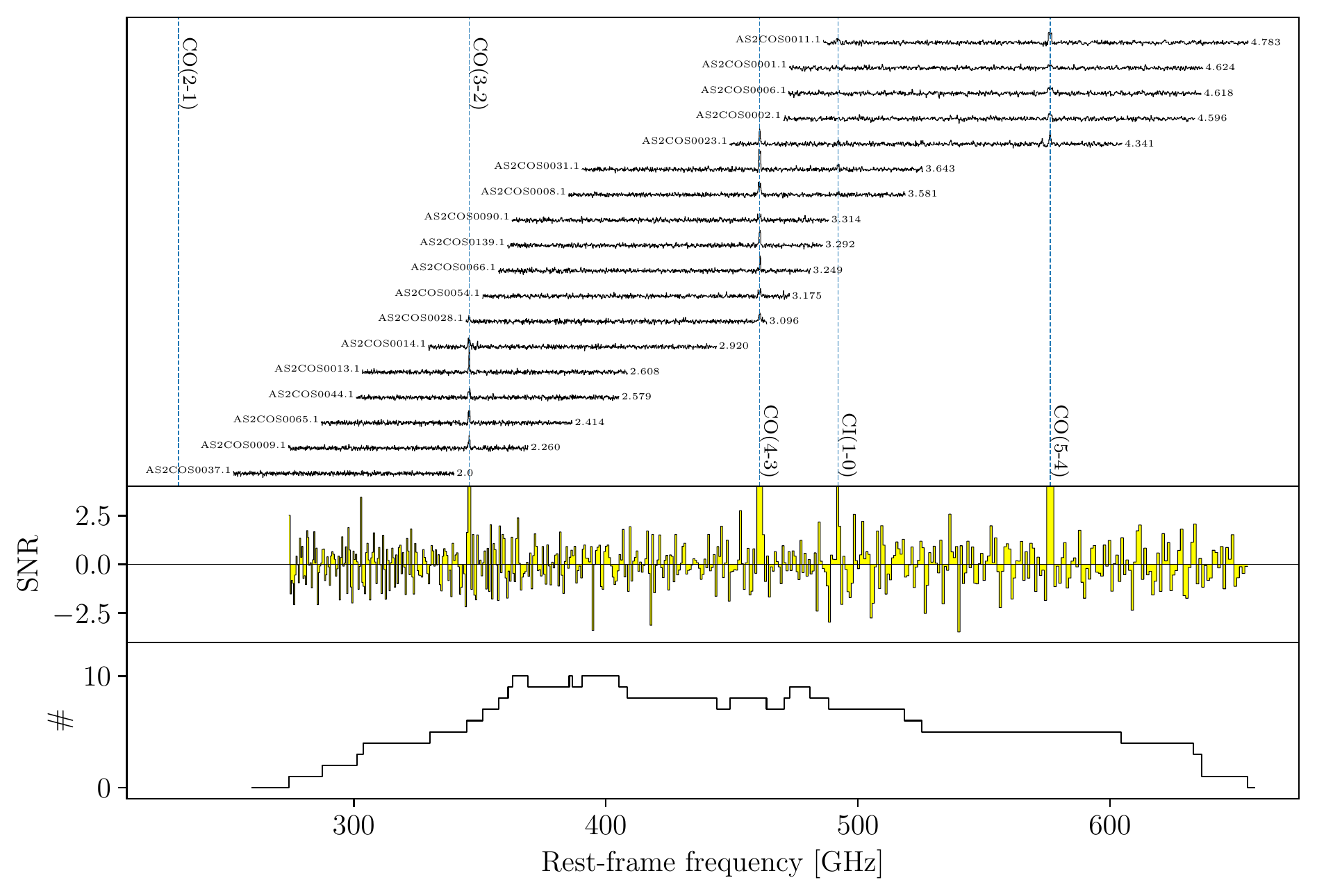}
		\caption{{\it Top:} An overview of the continuum-subtracted spectra of all our 18 primary SMGs in their rest-frame frequencies, along with the marked locations of strong molecular and atomic lines. Spectral lines are detected in 17 out of 18 targeted SMGs. SMGs with single line detection have their redshifts determined jointly with photometric redshifts, which can be shown to have $\sim$70-80\% success rate based on those that have multiple line detection (\autoref{sec:zdeter}; Birkin et al. 2021). AS2COS0037.1 is the only one SMG that is bright in optical and near-infrared but no lines detected in the scan, suggesting that based on its photometric redshift it is likely sitting within the known redshift gap for the band 3 scan. In this plot we therefore set the redshift of AS2COS0037.1 to 2.0 for demonstration purpose. {\it Middle:} Error-weighted stacked spectrum in signal-to-noise ratio, binned to 500\,km\,s$^{-1}$ per channel. The range in y-axis is from -4 to 4. AS2COS0037.1 is not included in the stack. No additional lines are detected in the stacked spectrum. {\it Bottom:} The number of spectra involved in the stack for each channel.
		}
		\label{fig:fig5}
	\end{center}
\end{figure*}

The spectra used for later analyses are extracted based on the coordinates presented in Simpson et al. (2020), which are based on ALMA 870\,$\mu$m continuum observations. All our targets have their 3\,mm continuum detected and we confirm that via 2D elliptical Gaussian fitting using {\sc imfit} the coordinates of the 3\,mm continuum best-fit peaks are consistent with those of 870\,$\mu$m continuum. We therefore adopt the 870\,$\mu$m coordinates given the significantly lower positional uncertainties. The results of {\sc imfit} also suggest that the 3\,mm continuum is not spatial resolved, which is expected since none strongly lensed SMGs like our targets are typically found to have sub-arcsecond sizes in rest-frame submillimeter (e.g., \citealt{Simpson:2015aa,Ikarashi:2015aa,Hodge:2016aa,Hodge:2019aa,Chen:2017aa,Fujimoto:2018aa,Gullberg:2019aa}). 

To determine the optimal aperture size for extracting the spectra and to derive the total line flux density, we employ the curve-of-growth analyses at the frequency of the emission lines and using the continuum subtracted cubes. For each source, spectra are extracted based on a set of apertures that are centered at the 870\,$\mu$m continuum position, shaped as a synthesized beam, and scaled by a range of factors in beam size from 0.5 to 4.5 with steps of 0.25. To estimate line flux density we conduct $\chi^2$ model fitting for each spectrum extracted, and two models, single Gaussian and double Gaussian, are considered. Given the channel-to-channel variations in sensitivity, the fittings are weighted by the noise of each spectral channel, which is estimated by taking the standard deviation of a collection of 1000 flux density measurements at random positions at same frequency but with the same aperture as that used to extract the target spectra. The line flux density of each extracted spectrum is then computed by integrating the best-fit model curves. 

\begin{figure*}[ht!]
	\begin{center}
		\leavevmode
		\includegraphics[scale=0.71]{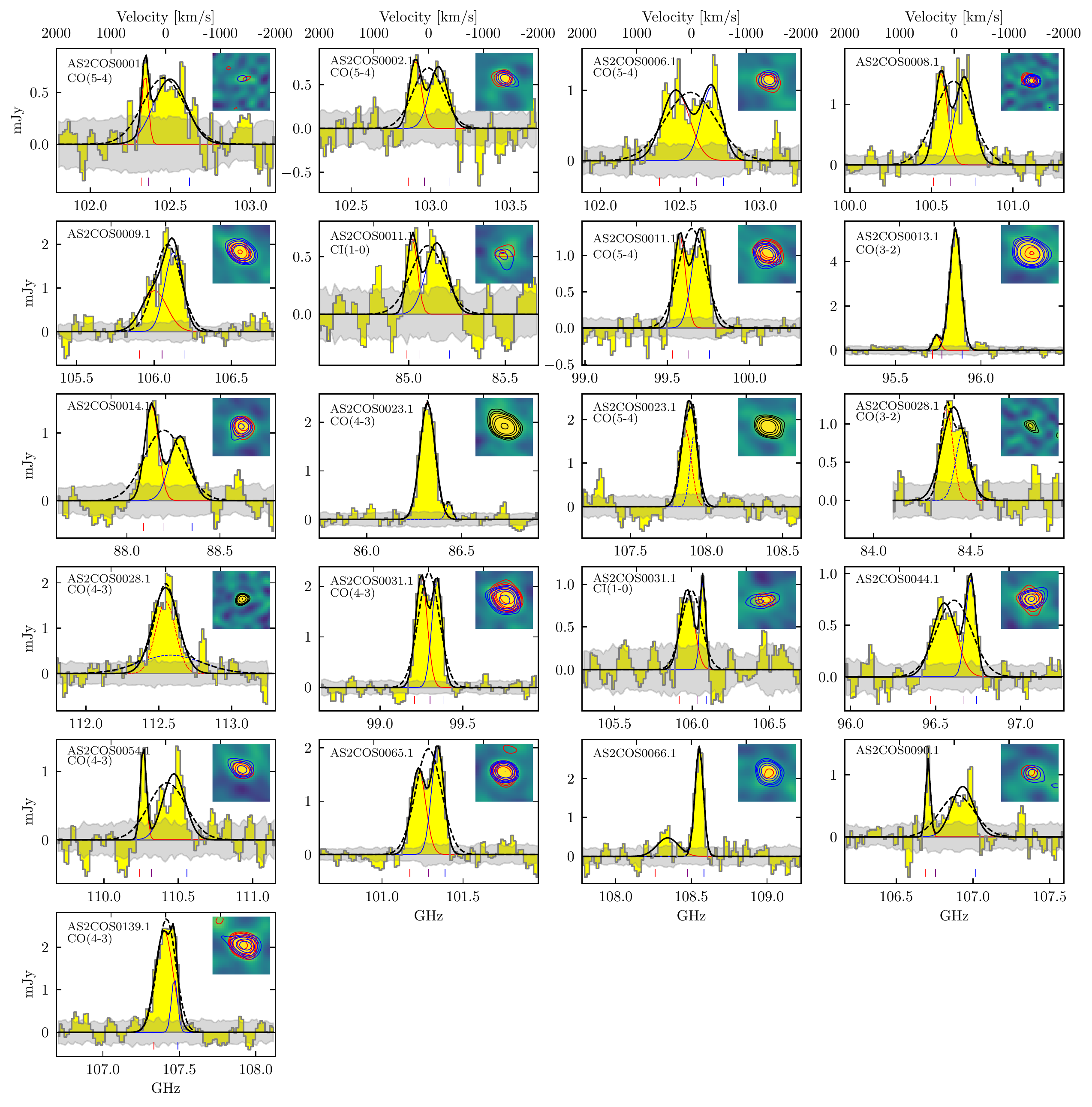}
		\caption{Spectra of all the detected line emissions from our primary SMGs, with their IDs indicated. All spectra are extracted using the aperture equivalent to the synthesized beam, and centered on the line centers of the best-fit models. The range of frequency plotted corresponds to a velocity range of $\pm$2000\,km\,s$^{-1}$. The best-fit single and double Gaussian models are plotted in dashed and solid curves, where the preferred models according to \autoref{table1} are plotted in solid. The red and blue components of the double Gaussian models are also shown. The gray bands indicate the r.m.s. levels. The transition of the line is also indicated if the redshift of the corresponding SMG is confirmed with more than one emission line. The inset of each spectrum shows the velocity-integrated emissions over the line FWHM, and it is shown in signal-to-noise ratio linearly scaled from -4 to 6 with a box size of 12.6$''$ on a side. The contours are plotted at levels of [3,4,5,7,9,15]\,$\sigma$; They are in black if single Gaussian model is preferred, or red and blue if double Gaussian model is selected. The small ticks below the lines that are better described by double Gaussian models mark the frequency ranges for making the velocity-integrated signals for the red and blue components.
		}
		\label{fig:fig6}
	\end{center}
\end{figure*}

Because of the two adopted Gaussian models the above approach results in two curve-of-growth curves for each target SMG. The selected examples are shown in \autoref{fig:fig4}, in which we can first observe that the integrated line flux densities are not sensitive to the different underlying models. Secondly, while some curves converge to a certain flux density level, as expected from a curve-of-growth analysis, some appear to be not converging. We suspect that this is due to larger-scale noise structures, which have also been shown to be the case in other ALMA data sets (e.g., \citealt{Novak:2020aa}). Indeed, by taking the median of all the curves (one from each target SMG) normalized at a given aperture size, one beam in this case, we find that this median curve nicely converges beyond around three times the beam size. In addition, this converging median curve is stable to within a few percent against the choice of aperture size to which all the curves are normalized, suggesting that this median curve can be used as a common profile for correcting the line integrated flux densities to total. Note that there is a hint of extended emissions by comparing the median curve to the curve-of-grown of the synthesized beam, although the difference is not significant (\autoref{fig:fig4}). The detailed size analyses will be presented in the forthcoming paper. However, to account for this possible extended emission we therefore adopt the median curve, instead of the curve of the synthesized beam for the correction to total (the difference is $\sim$20$\pm$10\%). We adopt the aperture equivalent to one synthesized beam partly because all the curves appear to be stable till this point and beyond which they start to diverge (\autoref{fig:fig4}). It also yields higher signal-to-noise ratios spectra compared to those extracted with larger aperture sizes, which help to determine a more appropriate model for the spectral profile (\autoref{sec:model}). 


Given the above justifications, the spectra that we use for analyses from this point on are therefore extracted using the aperture equivalent to the synthesized beam, and their estimated line flux densities are corrected to total based on the median of the curve-of-growth curves. For AS2COS0001.1, AS2COS0008.1, and AS2COS0028.1, where the nearby SMGs are also detected in a few cases and close enough to potentially affect the spectra, we re-image the cubes using the Briggs baseline weighting with a robust parameter of -1.5\footnote{We have also tried uniform weighting but some spectra become too noisy. The current setting represents a good balance between spatial resolution and signal to noise ratio.}, resulting into a $\sim$35\% reduction of the beam size, allowing clear separations to the companions with at least one synthesized beam distance ($\sim$3$''$). The spectra of these three SMGs are then extracted based on the smaller beam and again corrected to total using the corresponding median curve \footnote{The curve of growth analyses were re-run to all sources with the same weighting}. Finally for AS2COS0037.1, the only target SMG that we do not obtain significant line detection, its spectrum is extracted using the synthesized beam but no correction to total is applied. An overview of all the 18 spectra are shown in \autoref{fig:fig5}, where we also show the error-weighted stacked spectrum in signal-to-noise ratio. No additional emission lines are detected in the stacked spectrum. We have also experimented luminosity weighted stacking and luminosity normalized stacking, however the results remain unchanged.

\subsection{Model fitting}\label{sec:model}
The first analysis that we performed on the extracted spectra was to determine which model, single or double Gaussian, better describes the data. To make a quantitative assessment we employ both the Bayesian information criterion (BIC) and a version of Akaike information criterion that is modified for the limited sample size (AICc). The analyses are documented in \autoref{appendix:linefitting}, and the best-fit models are shown in \autoref{fig:fig6}. {In general we find that the results based on BIC and AICs are consistent to each other in 80\% of the cases, and they all point toward the fact that about 90\% of the primary SMGs have lines that are better described by double Gaussian profiles. The implications of this result are discussed later in the discussion section.}

Moment calculations are adopted to compute line properties using best-fit single and double Gaussian models, meaning
 \begin{equation*}
 	\begin{aligned}
 		&\bar{I} =  \int I_v dv \\
 		&\bar{v} = \frac{\int vI_v dv}{\int I_v dv} \\
 		&\bar{\sigma} = \frac{\int (v-\bar{v})^2I_v dv}{\int I_v dv}
 	\end{aligned}
 \end{equation*}
 and the first moment ($\bar{v}$), i.e., intensity-weighted central frequencies, are used to determine redshifts. The zeroth and the second moment will be used in the forthcoming paper to discuss line properties in detail.

 \begin{table*}[ht!]
	\caption{Redshifts of the primary target SMGs}
	\label{table2}      
	\centering
		\hspace*{-2cm}
		\begin{tabular}{lrlllll}
			\hline
			ID & $z_{\rm phot}$ & N$_{\text{L}}^a$ & Line & $z_{\rm spec}^b$ & $z_{\rm spec, adopted}^c$ & Other Lines\\
			\hline
			AS2COS0001.1 & 2.52+0.21-0.16 & ML & CO(5-4) & 4.6237$\pm$0.0007 & 4.6237$\pm$0.0007 & [CII]$^d$\\
			AS2COS0002.1 & 4.71+0.61-0.00 & ML & CO(5-4) & 4.5956$\pm$0.0006 & 4.5956$\pm$0.0006 & [CII]$^d$\\
			AS2COS0006.1 & 3.52+0.00-0.06 & ML & CO(5-4) & 4.6183$\pm$0.0001 & 4.6183$\pm$0.0001 & [CII]$^d$\\
			AS2COS0008.1 & 3.35+0.05-0.17 & SL & CO(4-3) & 3.5811$\pm$0.0004 & 3.5811$\pm$0.0004 & \\
			AS2COS0009.1 & 2.67+0.00-0.04 & SL & CO(3-2) & 2.2599$\pm$0.0002 & 2.2599$\pm$0.0002 & \\
			AS2COS0011.1 & 4.29+0.66-0.41 & ML & CI(1-0) & 4.7831$\pm$0.0007 & 4.7830$\pm$0.0002 & \\
			&  &  & CO(5-4) & 4.7830$\pm$0.0002 &  & \\
			AS2COS0013.1 & 2.46+0.02-0.05 & ML & CO(3-2) & 2.6079$\pm$0.0001 & 2.6079$\pm$0.0001 & CO(1-0)$^e$\\
			AS2COS0014.1 & 3.25+0.02-0.10 & SL & CO(3-2) & 2.9202$\pm$0.0005 & 2.9202$\pm$0.0005 & \\
			AS2COS0023.1 & 4.43+0.00-0.00 & ML & CO(4-3) & 4.3410$\pm$0.0001 & 4.3411$\pm$0.0001 & CO(1-0)$^e$ \\
			&  &  & CO(5-4) & 4.3414$\pm$0.0002 &  & \\
			AS2COS0028.1 & 3.40+0.06-0.17 & ML & CO(3-2) & 3.0966$\pm$0.0003 & 3.0965$\pm$0.0002 & \\
			&  &  & CO(4-3) & 3.0964$\pm$0.0002  & \\
			AS2COS0031.1 & 3.31+0.00-0.00 & ML & CO(4-3) & 3.6432$\pm$0.0001 & 3.6432$\pm$0.0001 & CO(1-0)$^e$ \\
			&  &  & CI(1-0) & 3.6431$\pm$0.0004 &  & \\
			AS2COS0037.1 & 2.35+0.04-0.02 & NL&  &  & 1.90$\pm$0.05 & \\
			AS2COS0044.1 & 2.98+0.14-0.05 & SL & CO(3-2) & 2.5793$\pm$0.0002 & 2.5793$\pm$0.0002 & \\
			AS2COS0054.1 & 2.73+0.00-0.00 & ML & CO(4-3) & 3.1755$\pm$0.0004 & 3.1755$\pm$0.0004 & CO(1-0)$^e$ \\
			AS2COS0065.1 & 2.40+0.29-0.01 & SL & CO(3-2) & 2.4140$\pm$0.0002 & 2.4140$\pm$0.0002 & \\
			AS2COS0066.1 & 2.90+0.06-0.06 & SL & CO(4-3) & 3.2492$\pm$0.0005 & 3.2492$\pm$0.0005 & \\
			AS2COS0090.1 & 3.48+0.85-0.11 & SL & CO(4-3) & 3.3137$\pm$0.0003 & 3.3137$\pm$0.0003 & \\
			AS2COS0139.1 & 2.23+0.00-0.02 & ML & CO(4-3) & 3.2923$\pm$0.0002 & 3.2923$\pm$0.0002 & Ly$\alpha^f$\\
			\hline
		\end{tabular}
		\begin{tabular}{l}
			Note: \\
			$^a$ Number of lines detected. ML stands for multiple lines, SL for single line, and NL for no line. \\
			$^b$ Redshift solutions for each line.
			$^c$ Redshift solutions for each source. 
			$^d$ \citet{Mitsuhashi:2021aa}  \\
			$^e$ Frias Castillo et al. in prep. \\
			$^f$ \citet{Daddi:2022aa}
		\end{tabular}
	\end{table*}

\subsection{Redshift determination}\label{sec:zdeter}

\begin{figure}[ht!]
	\begin{center}
		\leavevmode
		\includegraphics[scale=0.75]{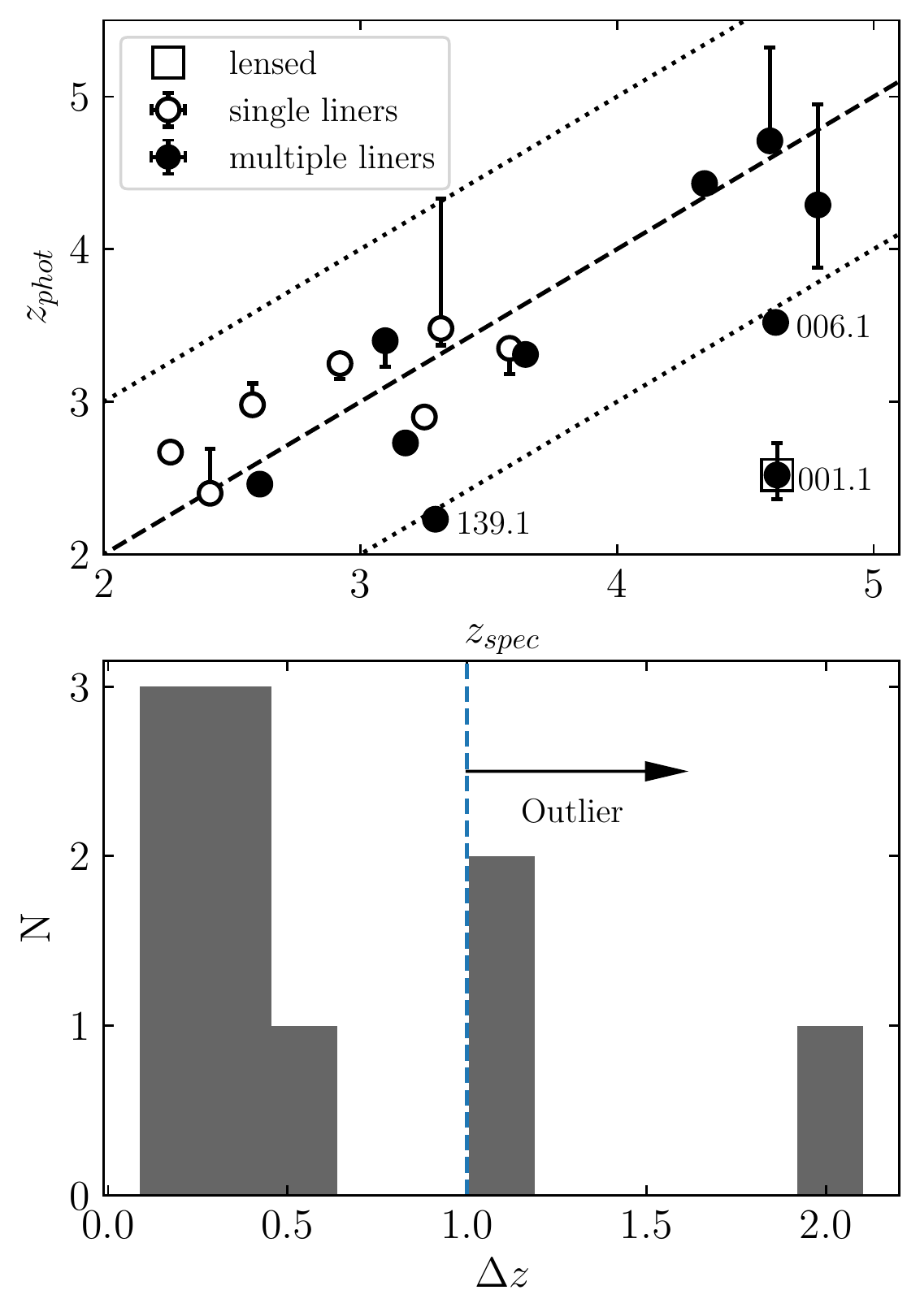}
		\caption{{\it Top:} Comparisons between photometric redshifts and spectroscopic redshifts on the 17 primary SMGs that are detected in lines, where the dashed line shows the one-to-one relation and the two dotted lines show $\Delta z\equiv \lvert z_{photo}-z_{spec}\rvert = 1$, outside of which is considered outlier. The outliers are defined as those that have their $\Delta z > 1$, meaning that if only single line detected it is assigned to an incorrect CO transition based on photometric redshifts. The ID numbers of the two outliers are indicated. The SMGs that have multiple lines detected are marked as filled circles and those with only one line detected marked as empty ones. The one that is boxed is AS2COS0001.1, due to the reason that its photometry could be contaminated by a nearby foreground lens galaxy. {\it Bottom:} Histogram of differences between photometric and spectroscopic redshifts ($\Delta z$), based on SMGs that have multiple lines detected. Excluding the two outliers, we find that the median redshift difference is $\langle\Delta z/(1+z_{\rm spec})\rangle=0.04\pm0.04$ with an bootstrapped error.
		}
		\label{fig:fig7}
	\end{center}
\end{figure}

To determine redshifts, the target SMGs are grouped into three categories; SMGs with multiple line detection, those with only one line detection, and then finally those that do not have emission line detection. Other work in the literature are referenced to aid the redshift determinations, in particular those from \citet{Jimenez-Andrade:2020aa}, \citet{Mitsuhashi:2021aa}, and \citet{Daddi:2022aa}. The category to which each SMG belongs is given in \autoref{table2}. In total, ten (56\%) SMGs have their redshifts determined by multiple emission lines, seven (39\%) by single emission line, and one (6\%) by its photometric redshift. 

For the SMGs with multiple line detection, the measured line frequencies are cross matched to combinations of redshifted emission lines, considering species that typically emit strongly at submillimeter such as carbon monoxide (CO) and atomic carbon ([CI]). We find an unique combination for each SMG and the results are listed in \autoref{table2}. The final adopted redshifts are determined in two ways; If the SMGs have multiple line detection in this reported ALMA data set, their redshifts are determined by taking a weighted average of the redshift solutions of each line. If the second line detection are originated from other work in the literature, we adopt our measurements since they tend to have higher precision. 

For the SMGs that only have one line detection, we find the solutions that are best matched to the photometric redshifts reported in Ikarashi et al. (2022), where as in \citet{Dudzeviciute:2020aa} the SED code {\sc magphys} is employed. We assume the line to be one of the CO transitions, which is justified by the fact that the frequency coverage of our data would have covered a CO line if the line is instead [CI]. Since in SMGs [CI] is typically weaker than the nearby CO lines ($J_{\rm up}=4,5$; \citealt{Spilker:2014aa,Birkin:2021aa}), if the line is [CI] then the nearby CO lines would have been detected. We note that compared to other methods, {\sc magphys} has the advantage of taking into account far-infrared and submillimeter photometry, allowing redshifts estimates for all AS2COSMOS SMGs, which are often times faint or undetected in optical and near-infrared wavebands where the known COSMOS catalogs such as COSMOS2015 \citep{Laigle:2016aa} and COSMOS2020 \citep{Weaver:2021aa} are based on. However given the rich information provided in the COSMOS catalogs and a good number of matches can be found on our target SMGs, we nevertheless exploit them later for detail assessments. 

Lastly, for the only SMG that has no line detection by far, AS2COS0037.1, the photometric redshift reported by Ikarashi et al. as well as the latest COSMOS2020 catalog ($1.93^{+0.04}_{-0.07}$ and $1.98^{+0.03}_{-0.15}$ based on the LePhare and EAZY codes, respectively; \citealt{Weaver:2021aa}) suggest that this SMG could be located within the known redshift gap of ALMA band 3 spectral scan, which is $z=1.74-2.05$. Indeed, given the depth reached by our data ($\sim$0.3\,mJy beam$^{-1}$), assuming a single Gaussian model of 3\,$\sigma$ peak and a typical line width of 500\,km\,s$^{-1}$, the line intensity limit is 0.5\,Jy\,km\,s$^{-1}$. Assuming the missing line is CO(3-2), the closest transition based on the photometric redshift, the line intensity limit corresponds to a limit on the CO(1-0) luminosity of $L'_{\rm CO(1-0)}=2-4\times10^{10}$\,K\,km\,s$^{-1}$\,pc$^2$, assuming a latest $r_{31}\equiv L'_{\rm CO(3-2)}/L'_{\rm CO(1-0)}=0.63$ \citep{Birkin:2021aa}. Given the infrared luminosity of $\sim10^{13}L_\odot$ (Liao et al. in prep.), this luminosity limit is well below the measured $L_{\rm IR}-L'_{\rm CO(1-0)}$ correlation and would have allowed detection of the line in $\gtrsim90$\% of the literature reported detection on SMGs \citep{Birkin:2021aa}. We therefore adopt 1.90$\pm$0.05 for AS2COS0037.1, which is simply the mean of the redshift gap with an uncertainty that makes the probability of it being outside the redshift gap equivalent to 3\,$\sigma$.

{In \autoref{table2} we report the redshift solutions of our sample on individual lines as well as the final adopted values for each SMGs based on the methods stated above. We adopt the redshift results based on the selected model for each SMG listed in \autoref{table1}, although we note that these results are not sensitive to the chosen single or double Gaussian models for the line fitting.}

In \autoref{fig:fig7} we compare the photometric redshifts and the adopted spectroscopic redshifts for all target SMGs except for AS2COS0037.1. We find that based on the ten SMGs that have multiple line detection so unambiguous redshifts, three sources have their spectroscopic redshifts differ from photometric redshifts by more than one, effectively meaning that if only single line detected it would have been assigned to an incorrect CO transition based on photometric redshifts. This suggests a 70\% success rate for our strategy of using photometric redshfits to aid determinations of spectroscopic redshifts for SMGs with single line detection. Although note that this success rate is estimated based on the higher redshift subset, which tends to have less reliable photometric redshifts. Thus the success rate for our sample SMGs that have one single line detection could be higher.

Our success rate is similar as that of a recent blind survey of emission lines on fainter SMGs \citep{Birkin:2021aa}, in which they found about 80\% success rate. Excluding the three outliers, we find that the median redshift difference is $\langle\Delta z/(1+z_{\rm spec})\rangle=0.04\pm0.04$ with an bootstrapped error, and a scatter of $\sigma_{\Delta z/(1+z_{\rm spec})}$ = 0.06, suggesting that the photometric redshifts estimated via SED fitting using {\sc magphys} can be accurate on average with good precision. However, we also find that the scatter is much larger than what can be inferred from the quoted uncertainties of the photometric redshifts, indicating that the uncertainties of photometric redshifts are underestimated. 

Based on the assessments above, we expect that for the seven SMGs that have only one line detection, about 1-2 sources may have their real redshifts different from, and likely higher than, the adopted values (details given in the \autoref{appendix:redshifts}). We take these into accounts when comparing our results to model predictions.



\section{Discussion} \label{sec:discussion}





\subsection{Physically associated pairs}
\begin{figure*}[ht!]
	\begin{center}
		\leavevmode
		\includegraphics[scale=0.85]{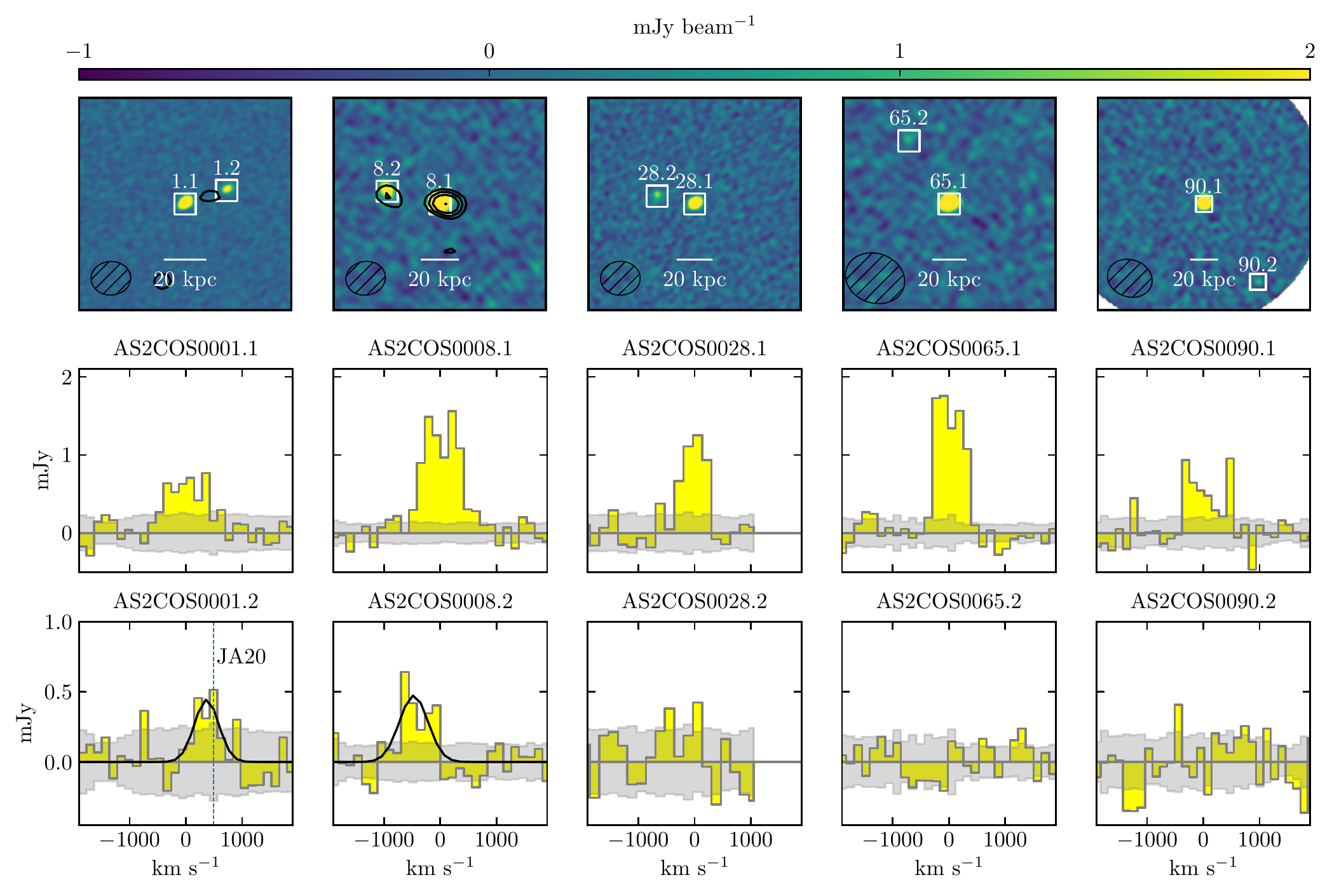}
		\caption{Continuum images at 870\,$\mu$m and 3\,mm spectra of the five sets of paired SMGs.  {\it Top:} Sub-arcsecond continuum images at 870\,$\mu$m from AS2COSMOS \citep{Simpson:2020aa}, with scale bars of 20\,kpc marked based on the adopted redshifts in \autoref{table2} for the primary SMGs. The images are uniformly and linearly scaled between -1 and 2\,mJy and the color bar plotted on top. The SMGs are boxes in white and their IDs are indicated. The synthesized beams of the 3\,mm spectra adopted for each pair are shown in the bottom left corners and the black contours are signal-to-noise ratios in 3, 4, 5, and 7\,$\sigma$ for the line emissions averaged over the FWHM based on the best-fit single Gaussian model profiles shown in the bottom spectra. {\it Middle and Bottom:} 3\,mm spectra binned to $\sim$150\,km\,s$^{-1}$ per channel for the primary ({\it middle panels}) and secondary ({\it bottom panels}) SMGs. These spectra are extracted using apertures centered on their reported 870\,$\mu$m locations (Simpson et al. 2020) with sizes and shapes of their corresponding synthesized beam. The velocities are referenced to the redshifts of the primary SMGs (\autoref{table2}), and the 1\,$\sigma$ r.m.s. levels are plotted in gray. For the secondary SMGs, we obtain significant detection from AS2COS0008.2, marginal detection from AS2COS0001.2, and no detection from AS2COS0028.2, AS2COS0065.2, and AS2COS0090.2. There is no line detection from AS2COS0028.2 in both frequency ranges covering CO(3-2) and CO(4-3), and for demonstration purposes we only show the range of CO(3-2). The spectra of AS2COS0001.2 and AS2COS0008.2 are fitted with single Gaussian model profiles and the best fits are plotted as black curves. The same line for AS2COS0001.2 was more significantly detected by \citet{Jimenez-Andrade:2020aa} using deeper NOEMA observations and the line peak deduced based on their results is shown in dashed blue line and marked as JA20. 
		}
		\label{fig:fig8}
	\end{center}
\end{figure*}

Interferometric follow-up observations in the past decade have convincingly shown that submillimeter sources detected by single-dish observations tend to break up into multiple sources, with a probability strongly correlated with the flux density of single-dish measurements, from $\sim$10\% for $\sim$3\,mJy SMGs to $>$50\% for SMGs at $\gtrsim$10\,mJy \citep{Wang:2011p9293,Barger:2012lr,Hodge:2013lr,Simpson:2015ab,Cowie:2018aa,Hill:2018aa,Stach:2019aa,Simpson:2020aa}. While the number of constituent SMGs can range from 2 to 4, it has also been shown that the flux densities measured from single-dish observations are dominated by the brightest constituent SMG, with a 70\% contribution on average. 

The relationship among these constituent SMGs is however unclear, in particular whether they are physically associated and undergoing dynamical interactions, or unrelated SMGs that are simply a result of line-of-sight chance projections. The determination of this relationship has profound implications in our understandings on the triggering mechanisms of star formation for SMGs. Observationally, statistical approaches using photometric redshifts and number density analyses have suggested that $\sim$30\% of these pairs are physically associated (e.g., \citealt{Stach:2018aa,Simpson:2020aa}), which is in agreement with spectroscopic measurements of small samples of paired SMGs \citep{Hayward:2018aa,Wardlow:2018aa}. Theoretically, models have instead predicted that the majority of these pairs are physically unassociated, especially for the bright submillimeter sources with single-dish measured flux density of $S_{\rm 850}\gtrsim10$\,mJy \citep{Hayward:2013qy,Cowley:2015aa,MunozArancibia:2015aa}. To definitely settle this issue spectroscopic follow-up observations are needed, and our sensitive spectral measurements on a complete sample of bright SMGs are suitable for providing further constraints.

Our sample include five pairs of SMGs and their 870\,$\mu$m continuum images and 3\,mm spectra are shown in \autoref{fig:fig8}. As stated in \autoref{sec:spec} and seen in \autoref{fig:fig8}, the spatial resolutions of the naturally weighted cubes are too coarse to separate the SMG pairs of AS2COS0001.1/1.2, AS2COS0008.1/8.2, and AS2COS0028.1/28.2. We therefore re-image those cubes using Briggs weighting with a robust parameter of -1.5, which stands as a good balance between sensitivity and resolution after experimenting different baseline weightings. To increase the detectability of line emissions we also binned the channels to $\sim$150\,km\,s$^{-1}$ per channel, appropriate for the typical linewidth of SMGs which is about 500\,km\,s$^{-1}$ in FWHM (e.g., \citealt{Bothwell:2013lp,Birkin:2021aa}). The spectra are extracted using apertures centered on their reported 870\,$\mu$m locations (Simpson et al. 2020) with sizes and shapes of their corresponding synthesized beam, shown at the corners of \autoref{fig:fig8}. 

We obtain a significant detection from AS2COS0008.2 and a marginal detection from AS2COS0001.2. The best-fit single Gaussian model suggests that the redshift of AS2COS0008.2 is 3.5739$\pm$0.0009 assuming that it is the same line as AS2COS0008.1, and the significance of the detection is at $>$4\,$\sigma$ based on the velocity-integrated intensity over the FWHM of the best-fit model (\autoref{fig:fig8}). Although marginal ($\sim2$\,$\sigma$ over the FWHM) in our data, the same line from AS2COS0001.2 (and AS2COS0001.1) was detected in high significance ($>5$\,$\sigma$) by \citet{Jimenez-Andrade:2020aa} with much deeper data from NOEMA, and \citet{Simpson:2020aa} and \citet{Mitsuhashi:2021aa} both reported detection of [CII] line from AS2COS0001.2. Their reported redshifts are consistent with our measurements. For the following analyses we therefore adopt the redshift measured by \citet{Jimenez-Andrade:2020aa} for AS2COS0001.2, which is 4.633$\pm$0.001.

The velocity differences between these two pairs are within 500\,km\,s$^{-1}$. At the projected distances of 20-30\,kpc, their velocity differences would suggest that these two pairs are gravitationally bound systems, assuming that SMGs as suggested by clustering analyses reside within halos with masses of $\sim$10$^{13}$\,$M_\odot$ \citep{Hickox:2012kk,Chen:2016ab,Wilkinson:2017aa,An:2019aa,Lim:2020aa,Stach:2021aa}. The redshift differences of the two pairs are both $\Delta z<0.02$, a definition adopted by some models for physically associated pairs \citep{Hayward:2013qy,MunozArancibia:2015aa}. This suggests that 40\% of the paired SMGs in our sample are physically associated. In addition, given that our observations are designed to detect lines from bright SMGs, deeper observations may reveal associated line emissions from the secondary SMGs of other pairs, and the true fraction of physically associated pairs could be higher. Indeed, as seen in \autoref{fig:fig8}, among these four secondary SMGs, those not detected in line by our data are the faintest in 870\,$\mu$m flux densities ($<$2\,mJy), and based on recent measurements \citep{Dudzeviciute:2020aa,Birkin:2021aa} we do not expect to detect lines on these faint SMGs given the sensitivity of our data. Our results are also consistent with \citet{Simpson:2020aa}, where based on the number counts analyses they estimated $62\pm7$\% of the $>3$\,mJy secondaries are physically associated with the primary SMGs.

While our results suggest a relatively high fraction ($\ge$40\%) of physically associated SMG pairs, the current sample size is admittedly too small to have much constraining power. Furthermore, it is also not straightforward to compare our results against model predictions. First of all, the adopted criteria in defining pairs are different in each model. For example, \citet{Cowley:2015aa}. made predictions based on the brightest two constituent SMGs. On the other hand, \citet{Hayward:2013qy} considered all SMGs with 850\,$\mu$m flux densities of $>$1\,mJy, which is the flux level that is highly incomplete in the parent AS2COSMOS catalog. Secondly, our spectral measurements are incomplete in terms of including all pairs or multiples of submillimeter sources above a certain flux density limit measured by SCUBA-2\footnote{For example, S2COS850.0003 is one of the brightest submillimeter sources in the S2COSMOS catalog \citep{Simpson:2019aa} but it breaks into four constituent SMGs in higher resolution ALMA maps and all of which have $S_{870}$ below our current flux cut \citep{Simpson:2020aa}.}, which is what model predictions are typically based on. To make fair comparisons, experiments that are tailored for these model predictions are needed.

In spite of these caveats, however, our results do provide constraints on this issue from another angle. Since our target SMGs are complete to their flux cuts, it can be stated that our results suggest that for SMGs brighter than 12.4\,mJy at 870 micron, their brightest companion SMGs are physically associated with the primary SMGs in $\ge$40\% of the time. 

\subsection{Redshift distribution} \label{sec:redshift}

Taking the adopted redshifts in \autoref{table2}, the overall median of the 18 primary target SMGs is 3.3$\pm$0.3 with a bootstrapping uncertainty, which is slightly larger than that of the photometric redshifts (3.1$\pm$0.2), simply due to the three outliers whose real redshifts are all higher than their photometric redshifts. Three target SMGs, AS2COS0001.1, AS2COS0002.1, and AS2COS0006.1, have been shown to be part of a $z=4.6$ coherent Mpc scale structure \citep{Mitsuhashi:2021aa}.

The uncertainty for the median caused by the measurement errors is on the order of $3\times10^{-4}$, which is the standard deviation of a set of 1000 medians, each of which is a result of a set of randomly perturbed redshifts generated according to the measurements. The uncertainty for the median caused by the wrongly assigned line transitions for the SMGs with single line detection is estimated to be 0.1, which is again the standard deviation of a set of 1000 medians, each of which is a result of a set of redshifts that are generated based on the adopted redshifts but having randomly chosen 2 SMGs with single line that are randomly assigned to one transition up or down from the adopted one, providing that the new transition would not allow the coverage of another CO transition within the bandwidth. In short summary, the dominant contributor to the uncertainty of the median is the bootstrapping uncertainty, essentially reflecting the sample size and the underlying distribution.


\begin{table*}
	\caption{Model catalogs used for comparisons}
	\label{table3}      
	\centering
	\begin{tabular}{llrll}
		\hline
		Model & Method & Area [degree$^{\text{2]}}$ & Lensing & References\\
		\hline
		Cai-Negrello & Analytical + Empirical & 500 & Yes & \citet{Cai:2013aa,Negrello:2017aa}\\
		SIDES & Empirical & 2 & Yes & \citet{Bethermin:2017aa}\\
		Casey-Zavala & Empirical & 10 & No & \citet{Casey:2018aa,Zavala:2021aa}\\
		Popping & Semi-empirical & 22.2 & No & \citet{Popping:2020aa}\\
		SHARK & Semi-analytical & 107.9 & No & \citet{Lagos:2019aa,Lagos:2020aa}\\
		\hline
	\end{tabular}
\end{table*}

\subsubsection{Comparisons with previous measurements}\label{sec:zvsobs}

\begin{figure}[ht!]
	\begin{center}
		\leavevmode
		\includegraphics[scale=0.8]{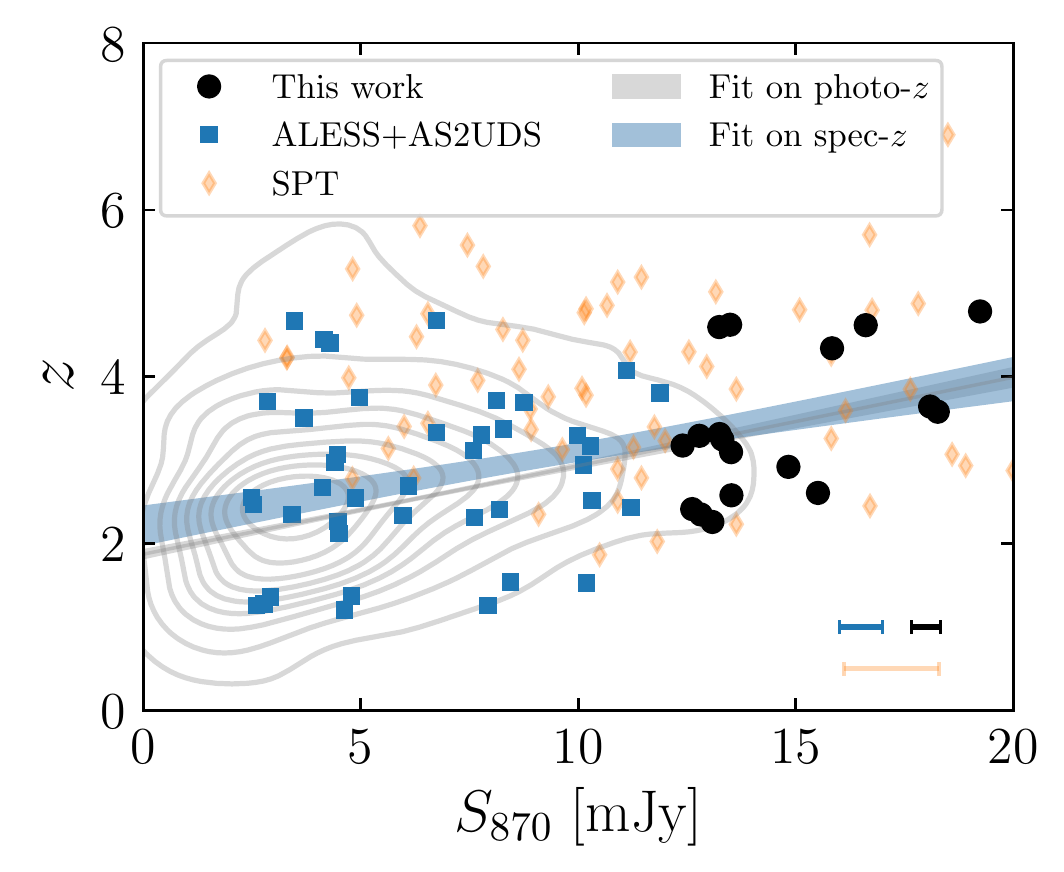}
		\caption{Redshift distribution as a function of $S_{870}$ based on uniformly selected SMG samples with ALMA follow-up observations. Spectroscopic redshifts of this work are plotted in black circles, and those for ALESS and AS2UDS subsamples are plotted in blue squares, which are reported by \citep{Birkin:2021aa}. We also plot the spectroscopic redshifts of the SPT SMGs versus their intrinsic flux densities \citep{Reuter:2020aa}. Typical uncertainties for the spectroscopic redshift samples are shown at the bottom right corner with matched colors. The background density contours show the distribution of photometric redshifts based on the three samples, ALESS \citep{Simpson:2014aa}, AS2UDS \citep{Stach:2019aa}, and AS2COSMOS (Ikarashi et al. 2022). The results of linear fittings on photometric redshifts and spectroscopic redshifts of the combined ALESS, AS2UDS, and our results are plotted as gray and blue bands, respectively, where the widths represent 68\% confidence levels. The slope of the best fit model on the spectroscopic redshifts is 0.09$\pm$0.02, consistent with that of photometric redshifts and previous studies \citep{Cowie:2017aa,Stach:2019aa,Simpson:2020aa,Birkin:2021aa}.
		}
		\label{fig:fig9}
	\end{center}
\end{figure}

Previously, redshift measurements of SMGs have been mostly focused on those with $S_{850/870}\sim2-10$\,mJy (e.g., \citealt{Chapman:2005p5778, Ivison:2007fj, Wardlow:2011qy, Simpson:2014aa, Cowie:2017aa, Stach:2019aa, Birkin:2021aa}). This is simply a result of the typical survey depth and area, coupled with the shape of the number counts, that maximizes the number of detection in this flux range. The median redshifts deduced differ slightly in different studies but they generally lie in the range of 2.4-2.6, with an uncertainty of about 0.1-0.2. Together with our results this hints that brighter SMGs are on average located at higher redshifts.

Since these previous studies used various methods to estimate or measure redshifts and they often covered different flux ranges. To better assess the correlation between flux density and redshift in \autoref{fig:fig9} we plot the most recent results from self-similar and well-defined samples, namely ALESS \citep{da-Cunha:2015aa}, AS2UDS \citep{Stach:2019aa}, and AS2COSMOS (\citealt{Simpson:2020aa}; Ikarashi et al. 2022). We investigate photometric and spectroscopic redshifts separately. In \autoref{fig:fig9} the photometric redshifts of all three samples are combined and shown as density contours. In both cases the correlation between flux density and redshift can be observed. Indeed the Spearman correlation coefficients are 0.3 and 0.4 for photometric  and spectroscopic redshifts, respectively, and the $p$-values are $\lesssim$0.001. The maximum likelihood linear fitting yields $z_{\rm median} = (2.2\pm0.3)+(0.09\pm0.02)\times S_{870}$ for the sample with spectroscopic redshift, consistent with the best-fit based on photometric redshifts. Our results are also consistent with previous assessments regarding either the general trends or the slope \citep{Ivison:2002uq,Cowie:2017aa,Stach:2019aa,Simpson:2020aa,Birkin:2021aa}. 

Finally, we compare our results to the SPT SMGs presented by \citet{Reuter:2020aa}, where complete measurements of spectroscopic redshifts are reported on a well-defined millimeter selected SMG sample. The reported SPT SMGs have apparent flux densites of $S_{870}>25$\,mJy and they are mostly strongly lensed. In \autoref{fig:fig9} we show their redshift distributions with respect to the intrinsic flux densities where gravitational magnifications are accounted for. Interestingly, the distribution of the SPT SMGs appears to follow our unlensed SMGs at $S_{870}\gtrsim10$\,mJy, below which the known selection effect of lensing biasing sources toward higher redshifts can be clearly seen.

\subsubsection{Comparisons with models}\label{sec:zvsmodel}
We now compare our measurements to model predictions. Currently there are various schools of models that can reproduce basic properties of SMGs such as number counts and have made predictions for other properties such as redshifts. They include empirical models (e.g., \citealt{Bethermin:2012fk,Bethermin:2017aa,Casey:2018aa,Zavala:2021aa}), semi-empirical models (e.g., \citealt{Hayward:2013lr,Popping:2020aa}), semi-analytical models (e.g., \citealt{Lacey:2016aa,Lagos:2020aa}), hybrid models including both analytical and empirical recipes (e.g., \citealt{Negrello:2007aa,Cai:2013aa,Negrello:2017aa}), and finally hydrodynamical simulations with various treatments of dust and its emissions (e.g., \citealt{McAlpine:2019aa,Hayward:2021aa,Lovell:2021aa}).

In general, while good progress has been made over the years, it is still difficult to find in hydrodynamical simulations as many bright SMGs as those detected in observations. For example, imposing the flux cut of our sample, there are only two such sources in the model presented by \citet{Lovell:2021aa}, and similar or fewer numbers are seen in other hydrodynamical models \citep{McAlpine:2019aa,Hayward:2021aa}. {The lack of SMGs brighter than our flux cut could be due to the fact that current simulations do not have sufficiently volumes, which are typically about 100 cMpc on a side.} As a result, for a meaningful comparison we do not consider predictions based on hydrodynamical simulations in this section.  

\begin{figure*}[ht!]
	\begin{center}
		\leavevmode
		\includegraphics[scale=0.5]{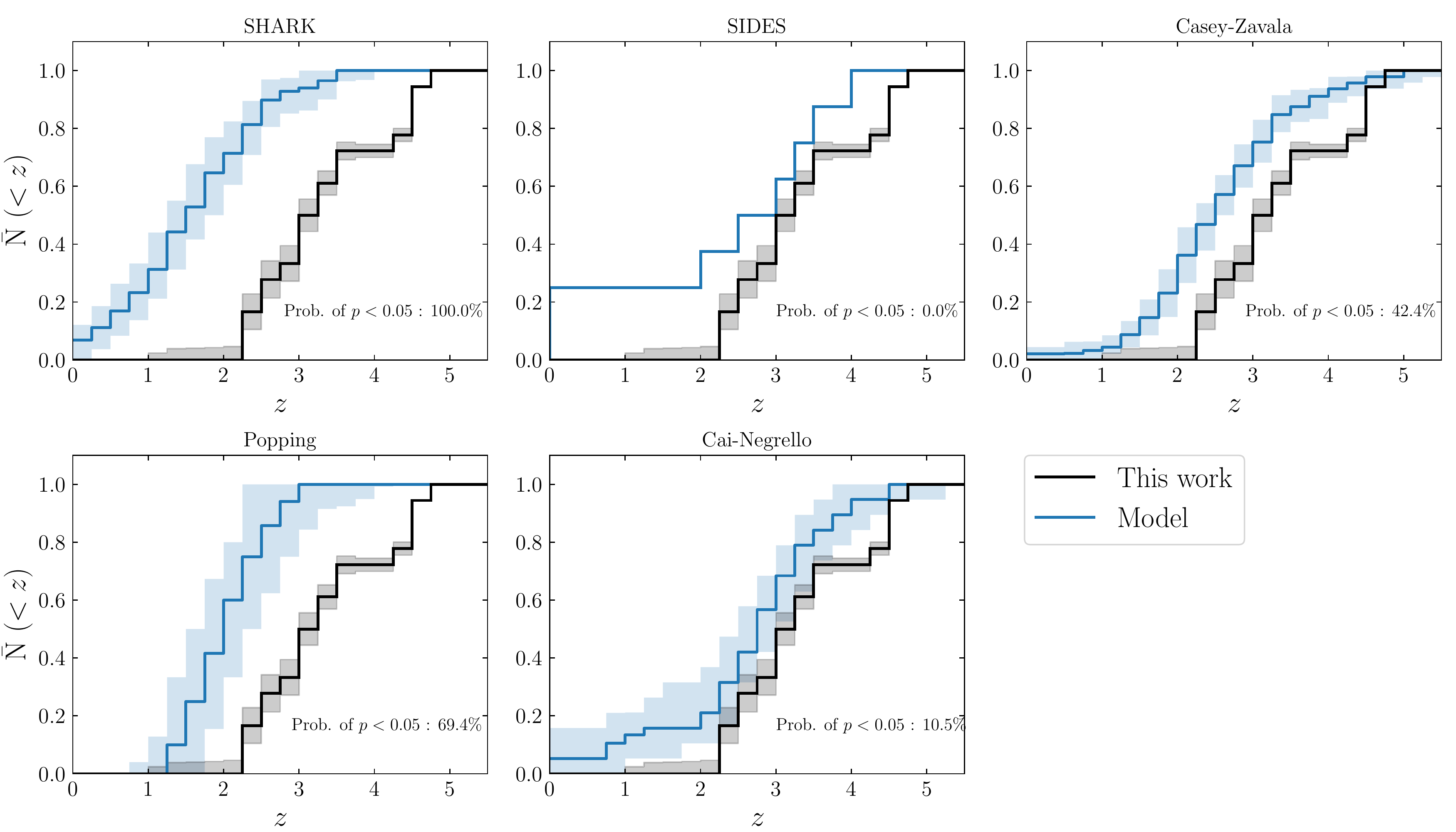}
		\caption{Cumulative distributions of our measured redshifts in black and the distributions of each model catalog, summarized in \autoref{table3}, are plotted in each panel in blue with their associated model names indicated on top. The 10-90 percentile distributions for each curves are based on simulations described in \autoref{sec:redshift}. In each panel, the probability of finding $p<0.05$ from a set of two-sample Kolmogorov–Smirnov tests based on simulations of the two comparison distributions are shown. 
		}
		\label{fig:fig10}
	\end{center}
\end{figure*}

We list the model catalogs used for comparisons and their details in \autoref{table3}, including the area from which the catalogs are drawn and the methods used to generate these catalogs. {Briefly, all these models employ certain methods to predict SFRs thus total infrared luminosity ($L_{\rm IR}$), and adopt various templates or relations to model the infrared and submillimeter SEDs. For example, analytical solutions of SFRs are calculated in physically motivated models like SHARK and Cai-Negrello, where the predicted submillimeter flux densities are modeled using  selected SED templates. For models like SIDES and Popping, SFRs are estimated by fitting their dark matter base empirical models to stellar mass functions with considerations of the star formation main sequence. SIDES then adopt SED template to model submillimeter fluxes, where Popping apply a relation between the 850 micron flux density and SFRs and dust masses based on dust radiative transfer codes. Casey-Zavala obtains $L_{\rm IR}$ directly by modeling the infrared luminosity functions, and a modified black body plus mid-infrared power-law SED model was assumed together with a relation between dust temperature and $L_{\rm IR}$. The full descriptions on the methodologies used to generate these model catalogs can be found in the latest reference for each model listed in \autoref{table3}. Although note that for Cai-Negrello the model has been slightly updated with a new version of SED template, resulting in a better agreement with the recently reported number counts in far-infrared and submilllimeter (Negrello et al. in preparation).} 

We assess their redshift distributions as the following. For SIDES since it is drawn from an area that has a size similar to our initial SCUBA-2 survey, we simply impose the flux cut of our sample and plot the cumulative distribution in \autoref{fig:fig10}. For model catalogs that are drawn from areas that are much larger than 2 square degrees, we randomly draw the model catalogs 100 times based on the size of our surveyed area. For each draw we assess their cumulative distribution and in \autoref{fig:fig10} we plot the median of all the 100 draws as well as their 10-90 percentile range to roughly indicate the fluctuations of these distributions mainly due to their field-to-field variations and {shot noise}.

To compare our observational results to models. We generate a set of 100 randomly perturbed redshift distributions based on our measurements and their associated uncertainties, as well as the probability of wrongly assigned redshifts for SMGs with only single line detection, following the procedure described in the previous paragraph. The results are plotted in \autoref{fig:fig10} again with 10-90 percentile range. We then perform two-sample Kolmogorov–Smirnov (KS) test between each of the 100 perturbed observational distributions and each of those 100 drawn from each of the model catalogs, so a total of 10000 comparisons except for SIDES which only has 100. For each model catalog, we record the $p$-values of each comparison and compute the percentage of them having $p<0.05$, a standard value that suggests the difference between the two samples is highly significant. Changing this $p$-value cut does not affect the conclusion. We show the results of these comparisons in \autoref{fig:fig10}. {Strikingly all models predict lower redshifts compared to our measurements, although from the statistical point of view the significance of their differences differ. Our measurements agree best with SIDES, however SIDES only provides one sight line. With a much larger volume perhaps the comparison against SIDES would yield a result that is similar to that from the comparison against Cai-Negrello. The most significant difference comes from the comparison against SHARK, where all sight lines that we draw do not agree with our measurements.}

We can also compare them in {median redshifts, which} are 1.7$\pm$0.3 for SHARK, 2.9$\pm$0.7 for SIDES, 2.6$\pm$0.3 for Casey-Zavala, 2.1$\pm$0.4 for Popping, and 2.9$\pm$0.4 for Cai-Negrello. The error budget again includes both statistical and systematic uncertainties caused by field-to-field variations and the shot noise of simulations. Our measured median redshift of 3.3$\pm$0.3 is larger compared to all models, with the least agreement to SHARK, which is in 3.8\,$\sigma$ tension. These are the same conclusions we make based on more detail two-sample KS tests. Measurements based on a much larger sample size (i.e., a factor of few more) would allow meaningful tests on the surviving models. 

\subsubsection{Implications}\label{sec:zimplication}

The physical reasons for brighter SMGs lying at higher redshifts remain unclear. It is worth noting that among models that are used to compare with our results, SHARK is the only one that is created with a complete suite of physically motivated treatments including merger trees of cold dark matter and baryon physics. It is also intriguing to note that another self-consistent and physically motivated semi-analytical model, GALFORM \citep{Lacey:2016aa}, made with different treatment of physics, has also predicted a low redshift ($z<2$) for the flux range of our sample and also a negative correlation between $S_{850/870}$ and redshift  \citep{Cowley:2015aa}.

Other physically motivated models based on hydrodynamical simulations like EAGLE \citep{McAlpine:2019aa}, Illustris/IllustrisTNG \citep{Hayward:2021aa}, and SIMBA \citep{Lovell:2021aa} have recently made good progress in matching number counts at the typical flux regime ($\lesssim$10\,mJy). While they are yet to be capable of making significant amount of $\gtrsim10$\,mJy SMGs, {in part due to the limited volume,} many are successful in reproducing median redshifts of fainter SMGs, although not all can reproduce the trend between flux density and redshift \citep{McAlpine:2019aa}. These efforts have pointed to the need to better understand the properties of dust, {in particular the relationship between dust, metal, and gas masses such as the dust-to-metals ratios, in order to address the issue that models tend to under predict dust masses for SMGs.} For example recent measurements of dust mass of SMGs have arrived to an average of about $6-7\times10^8$\,$M_\odot$ \citep{Dudzeviciute:2020aa,da-Cunha:2021aa}, somewhat higher than what is seen in most models. In our forthcoming paper (Liao et al. in preparation) we will show that our sample SMGs have even higher dust masses. On the other hand, sub-grid physics regarding the feedback processes has been suggested to be the key in making enough amount of dust in models \citep{Hayward:2021aa}. 

{Moving forward, observational constraints on various physical properties would be required to solve the puzzle of the formation of SMGs. For example, measuring dust-to-metal ratios for dusty galaxies through observations would be key in testing the predicted values from recent models. ALMA has been proven invaluable on the measurements of dust and gas masses, as well as dust properties such as dust emissivity index and temperature. Measurements of gas-phase metalicity, on the other hand, has been a challenging task for dusty galaxies due to them often being faint in rest-frame optical. One can always start with those that are significantly detected on the optical strong emission lines, although one needs to also bear in mind the possible biases toward the galaxies that may have the largest spatial offsets between dust and optical emissions, or the subset that is the least obscured. Far-infrared fine structure lines from [N\,{\sc ii}] and [O\,{\sc iii}] offer a promising way to measure gas-phase metalicity of dusty galaxies (e.g., \citealt{Rigopoulou:2018aa}), but a future facility such as a sensitive far-infrared probe satellite would be required to conduct such measurements on a large sample at cosmic noon. Finally, upcoming facilities such as JWST could provide sub-kpc rest-frame near-infrared imaging, which would help constrain stellar mass and morphology, addressing the issue of the triggering of star formation which has also been a topic of debate.}


\subsection{Gravitational lensing}\label{sec:lensing}
Strong gravitational lensing (lensing magnifications $\mu>2$) is found to be responsible for the exceptional apparent brightness ($S_{500\,\mu m}\gtrsim100$\,mJy or $S_{1.4\,mm}\gtrsim10$\,mJy or $S_{850\,\mu m}\gtrsim30$\,mJy) of the majority of the submillimeter and milllimeter sources that are uncovered by shallow but wide surveys such as those conducted by the SPT and {\it Herschel} \citep{Negrello:2010lr,Negrello:2017aa,Bussmann:2013aa,Hezaveh:2013aa,Wardlow:2013aa,Spilker:2016aa}. Most models predict that the fraction of sources experiencing strong lensing drops significantly, from about $\gtrsim$70\% to almost zero, within a narrow range of flux density, which is about 80-100\,mJy at 500\,$\mu$m and 10-30\,mJy at 850\,$\mu$m \citep{Wardlow:2013aa,Negrello:2017aa}. Our statistical sample SMGs have their flux densities within this range and therefore represent an opportunity to test model predictions.

\begin{figure*}[ht!]
	\begin{center}
		\leavevmode
		\includegraphics[scale=0.98]{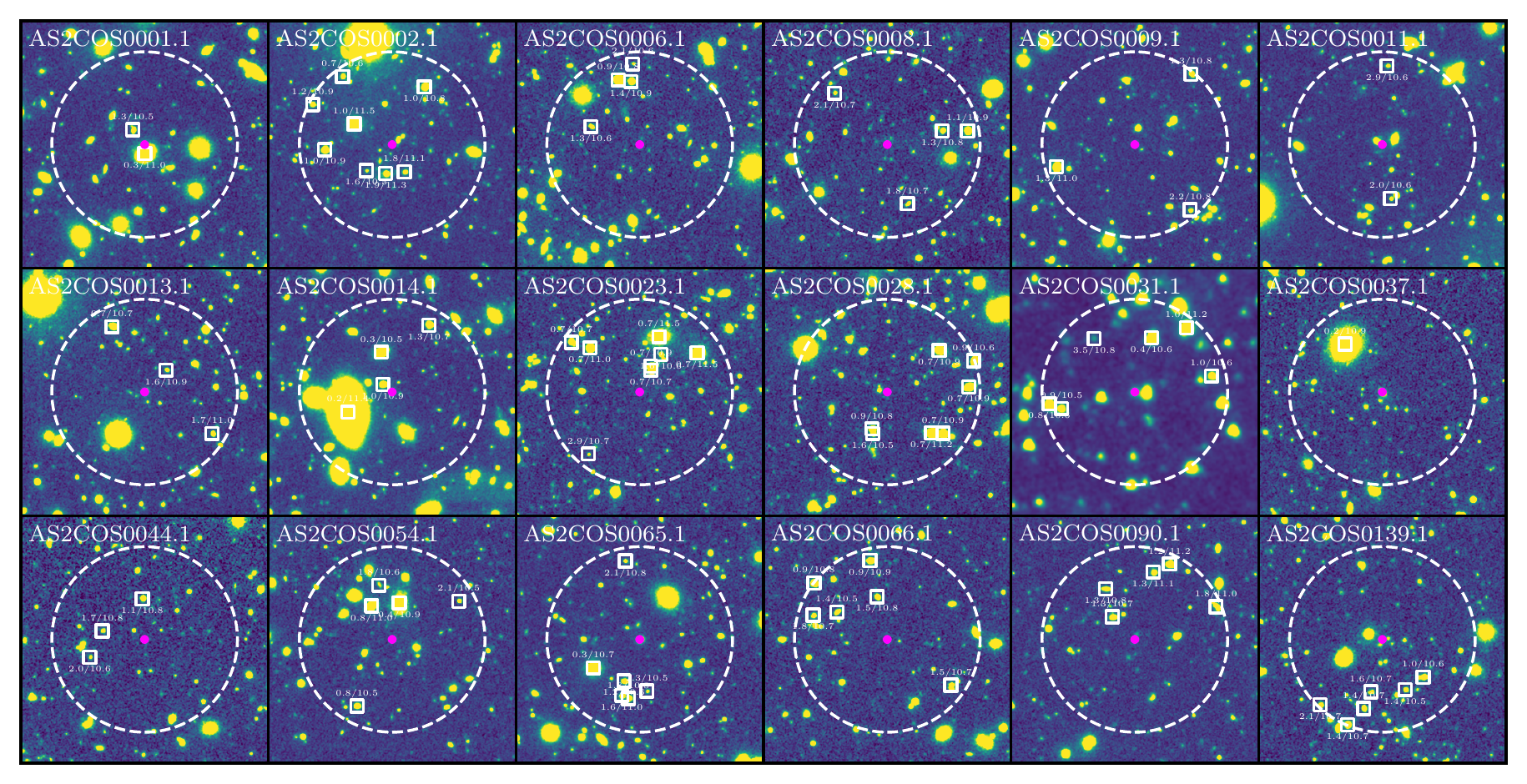}
		\caption{Thumbnail images in $K$-band (3.6\,$\mu$m for AS2COS0031.1) of our 18 sample SMGs. The large dashed white circles with a radius of 30$''$ and centered at our sample SMGs (magenta points) show the search area for constructing the foreground potential. Small unfilled boxes mark the neighboring foreground galaxies that have stellar masses $>10^{10.5}$\,$M_\odot$. The corresponding small values for each box are redshift in the left and stellar mass in logarithm base 10 in the right.
		}
		\label{fig:fig11}
	\end{center}
\end{figure*}

To estimate lensing magnifications of our sample SMGs we employ {\sc lenstool} \citep{Kneib:1996p3751,Jullo:2009aa}. For each SMG, we utilize the COSMOS15 catalog \citep{Laigle:2016aa} and construct a list of neighboring galaxies that are located within 30$''$ of the SMG and have had reported estimates of photometric redshifts and stellar masses. This angular distance is chosen such that beyond which the results do not change significantly. A gallery of the selected massive subsets of these neighboring galaxies are plotted for each SMG in \autoref{fig:fig11}. To construct a model of foreground gravitational potential for each SMG, we assume a regular NFW density profile ($\alpha=1$; \citealt{Navarro:1997aa}) for each neighboring galaxy in the list. Besides sky positions, other key input parameters for the model of each neighboring galaxy include ellipticity, positional angle, scale radius $r_s$, and characteristic velocity $\sigma_s$ \citep{Limousin:2005aa}. We describe the procedures in obtaining these parameters in the following.

We first run {\sc SExtractor} \citep{Bertin:1996zr} to obtain their structural parameters ellipticity and positional angle, and the primary image considered is the K-band image from UltraVista DR4, except for AS2COS0031.1 where it is outside the UltraVista footprint so we adopt the 3.6\,$\mu$m image from {\it Spitzer}. The scale radius and characteristic velocity are related to the assumed NFW profile, which is expressed as 
\begin{equation*}
\rho(r) = \frac{\rho_s}{(r/r_s)(1+r/r_s)^2}
\end{equation*}
where $r_s$ is the scale radius and $\rho_s$ is the characteristic density so that the characteristic velocity is $\sigma_s=4/3Gr_s^2\rho_s$. $\rho_s$ can be related to critical density $\rho_c$ as $\rho_s = \delta_c\rho_c$, where the density contrast $\delta_c = 200c^3/(3{\rm ln}(1+c)-3(c/1+c))$. $c$ is the concentration factor defined as the ratio between the halo radius $r_{200}$ (the mean density within which is 200$\rho_c$) and $r_s$. The halo mass can therefore be deduced as $M_{200}=800/3\pi r_{200}^3\rho_c$. That is, $r_s$ and $\sigma_s$ can be derived for a given combination of concentration factor and halo mass. 

For each neighboring galaxy, we estimate its halo mass by adopting the stellar mass and redshift provided by COSMOS15 catalog and applying the latest stellar-to-halo mass relationships given by \citet{Legrand:2019aa}. For concentration we refer to the classical work of \citet{Wechsler:2002aa} based on dark matter N-body simulations of $\Lambda$CDM cosmology, where the relationships between concentration, halo mass, and redshift are shown. 

Each of the adopted input parameters has its own reported uncertainty or scatter, and they should be taken into account in order to properly assess the probability distributions of lensing magnifications. To do so, for each SMG we create 100 potential models where each constituent potential is created by randomly perturbing each of their parameters based on their quoted uncertainties. We run {\sc lenstool} on these 100 potential models for each SMG and calculate the median and 1\,$\sigma$ percentile. The results are given in \autoref{table4}. 

As a check, the lensing magnification of AS2COS0001.1 was also estimated by \citet{Jimenez-Andrade:2020aa} with a different assumption of the density profile, and they found a factor of 1.5, consistent with our results. Notably, we find that if we only look at the mass distributions of a much closer surroundings, the magnifications of a few SMGs, such as AS2COS0002.1 and AS2COS0014.1, would be significantly underestimated. This is because in these cases there are a few very massive (up to $10^{11.5}$\,$M_\odot$) foreground galaxies, and in the case of AS2COS0002.1 a number of galaxies including the very massive ones are located at $z\sim1$, suggesting a group environment.

\begin{table}
	\caption{Lensing magnifications}
	\label{table4}      
	\centering
		\begin{tabular}{ll}
			\hline
			ID & $\mu$\\
			\hline
			AS2COS0001.1 & 1.4$^{+0.1}_{-0.1}$\\
			AS2COS0002.1 & 3.0$^{+1.4}_{-0.7}$\\
			AS2COS0006.1 & 1.1\\
			AS2COS0008.1 & 1.1\\
			AS2COS0009.1 & 1.0\\
			AS2COS0011.1 & 1.1\\
			AS2COS0013.1 & 1.1\\
			AS2COS0014.1 & 1.7$^{+0.5}_{-0.2}$\\
			AS2COS0023.1 & 1.7$^{+0.1}_{-0.1}$\\
			AS2COS0028.1 & 1.1\\
			AS2COS0031.1 & 1.1\\
			AS2COS0037.1 & 1.1\\
			AS2COS0044.1 & 1.1\\
			AS2COS0054.1 & 1.1\\
			AS2COS0065.1 & 1.0\\
			AS2COS0066.1 & 1.1\\
			AS2COS0090.1 & 1.1\\
			AS2COS0139.1 & 1.1\\
			\hline
		\end{tabular}
        \begin{tabular}{l}
			Note: Details can be found in \autoref{sec:lensing}. \\
			Uncertainties less than 0.05 are omitted.
		\end{tabular}    	
\end{table}

\begin{figure*}[ht!]
	\begin{center}
		\leavevmode
		\includegraphics[scale=0.65]{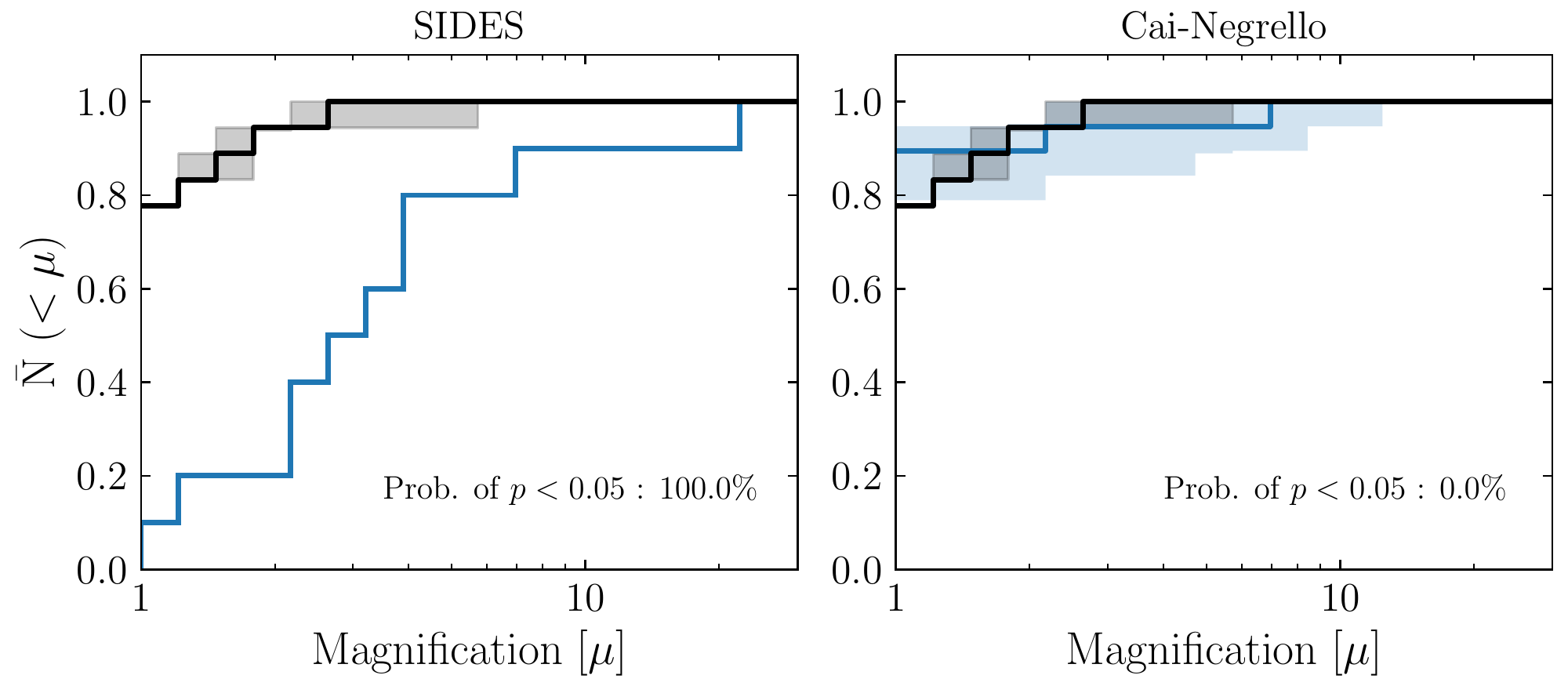}
		\caption{Cumulative distributions of estimated lensing magnifications in black and those based on model predictions of SIDES and Cai-Negrello are plotted in blue with their corresponding model names indicated on top. The 10-90 percentile distributions for each curves are based on simulations done in a similar fashion as that described in \autoref{sec:redshift}. In each panel, the probability of finding $p<0.05$ from a set of two-sample Kolmogorov–Smirnov tests based on simulations of the two comparison distributions are shown. 
		}
		\label{fig:fig12}
	\end{center}
\end{figure*}

Based on \autoref{table4}, we can see that by looking at their median only one out of the 18 SMGs in our sample can be classified as being strongly lensed with $\mu>2$, suggesting a relatively low strongly lensed fraction of $<10\%$ compared with the brighter sources. This is in line with what has been predicted by some observations \citep{Chapman:2002ab} and models. We note that our methodology in constructing foreground gravitational potentials does not account for contributions from halos larger than galactic halos (i.e., group scale halos) thus it is possible that  magnificantions of some SMGs may have been underestimated by a moderate amount. On the other hand, our methodology also does not account for under-dense regions so the overall magnification is slightly biased high, and as a result many unlensed SMGs have $\mu=1.1$. Nevertheless, these two factors are expected to not affect our results significantly.

We compare our results with predictions of magnification for the flux range of our sample from SIDES and Cai-Negrello in \autoref{fig:fig12} and find that our results are inconsistent with SIDES but consistent with Cai-Negrello. This can be understood by the fact that while the Cai-Negrello model takes into account both the redshift distribution of model SMGs and the mass distributions in the foreground, the predictions from SIDES are simply drawn from some probability distributions according to the redshifts of the model SMGs, as stated in Bethermin et al. (2017). 

\begin{figure}[h]
	\begin{center}
		\leavevmode
		\includegraphics[scale=0.75]{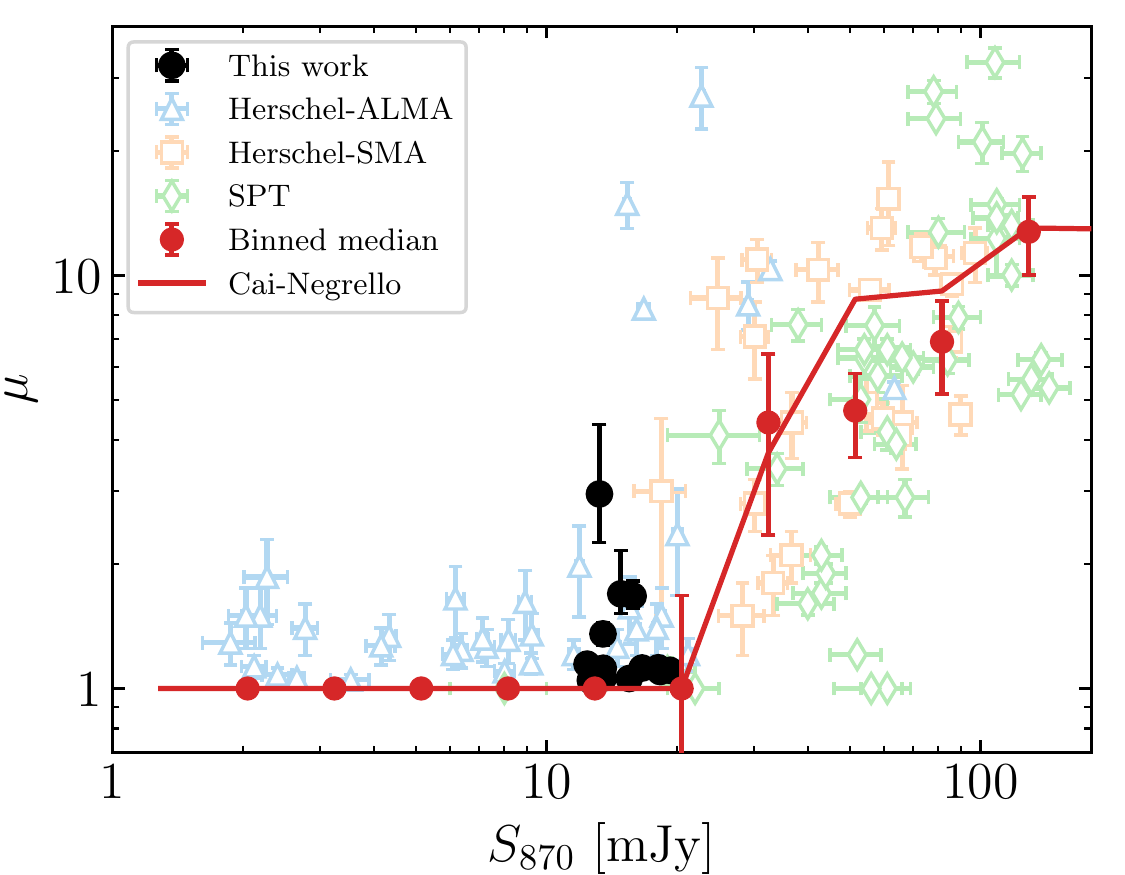}
		\caption{Lensing magnification as a function of 870\,$\mu$m flux density. The estimates for our sample SMGs are shown in black circles, and various estimates from other studies are plotted in different styles; blue triangles for a ALMA study of Herschel selected sources \citep{Bussmann:2015aa}, orange squares for a SMA study of a brighter Herschel-selected sample (Bussmann et al. 2013), and green diamonds show the results from the SPT survey \citep{Spilker:2016aa}. The red circles are binned medians where magnifications of $\mu<2$ are assigned to 1, following the method that is used to generate the model catalog by Cai-Negrello (Cai et al. 2013; Negrello et al. 2017), which is shown as the red curve. Overall we find the measurements are consistent with model predictions across this two orders of magnitude flux range.
		}
		\label{fig:fig13}
	\end{center}
\end{figure}

\begin{figure}[h]
	\begin{center}
		\leavevmode
		\includegraphics[scale=0.65]{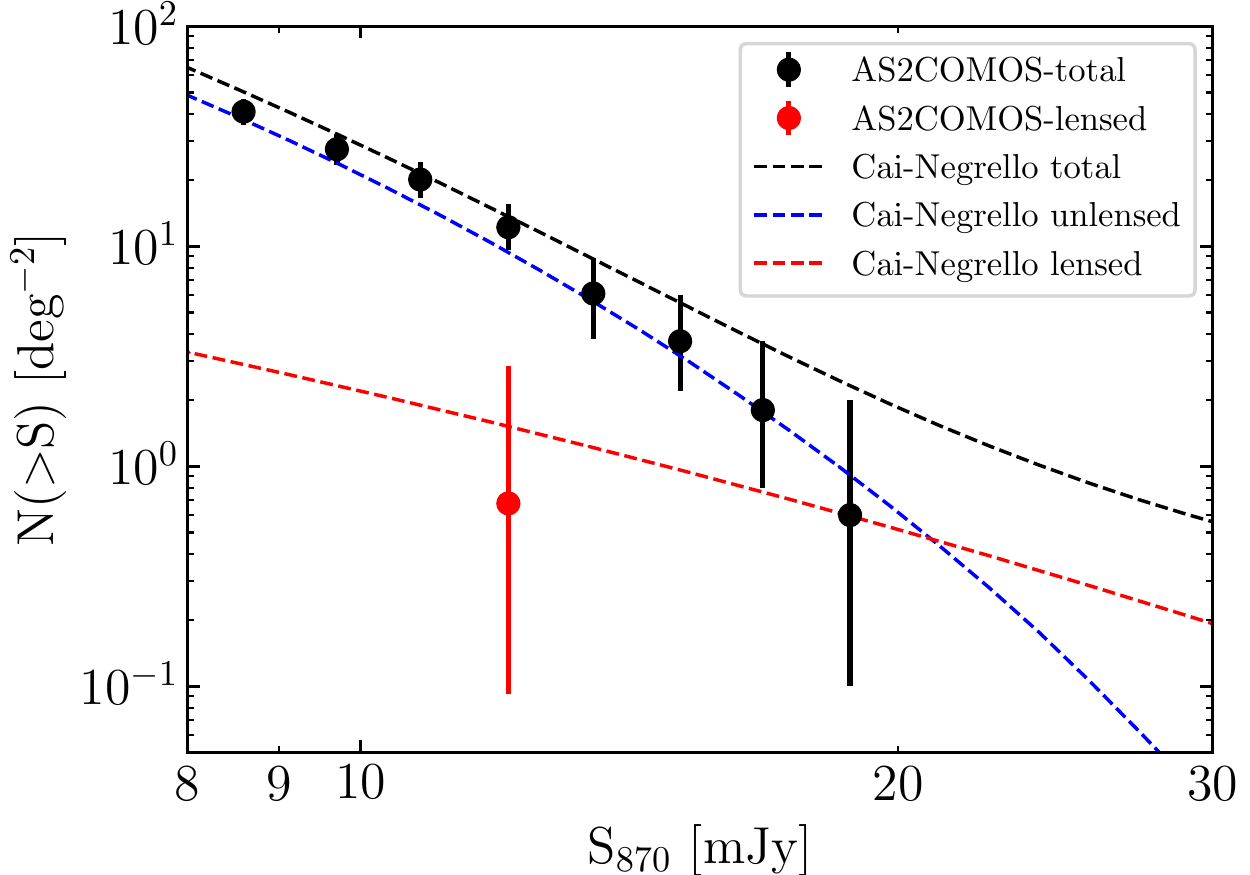}
		\caption{Cumulative number counts of the parent AS2COSMOS sample in black circles, and the counts of strongly lensed ($\mu>2$) SMGs are plotted in red circles, which are scaled from the counts of the parent sample with a factor of 1/18, the strongly lensed fraction based on \autoref{table4}. The counts predictions based on Cai-Negrello model are plotted in black for total counts, blue for none strongly lensed counts, and red for strongly lensed counts. 
		}
		\label{fig:fig14}
	\end{center}
\end{figure}

We now expand the comparisons by including published measurements for a wide range of 870\,$\mu$m flux density, including the SMA and ALMA follow-up studies of Herschel-selected sources \citep{Bussmann:2013aa,Bussmann:2015aa} as well as those based on the SPT survey (Spilker et al. 2016). For each of the SPT sources separate magnification factors were reported for each of their constituent components, so we compute their flux-weighted magnification as the nominal magnification for each SPT source (Figure 3 in \citet{Spilker:2016aa}). It is also worth noting that while the Herschel-selected sample observed by the SMA is a flux limited sample, the one observed by ALMA was selected to have a nearby foreground source to maximize the probability of detecting strongly lensed objects. As a result the Herschel-selected sample observed by ALMA is likely biased high in lensing magnification for the same flux range. 

The results are plotted in \autoref{fig:fig13}, where we find that despite the possibility of having a bias, the Herschel-ALMA measurements are consistent with our results at the overlapping flux density range (10-20\,mJy). We can also compare to the model predictions from Cai-Negrello. Note that in the Cai-Negrello model sources that are not strongly lensed ($\mu<2$) are assigned 1 for their magnification. We therefore adopt the same method for the measurements and compute the binned median with bootstrapped errors. We compare our results with the median of the model prediction with the same flux bins and we find they have good agreement overall.

Last but not least, having lensing magnification measurements of a flux limited sample means that we can compute the number counts of the strongly lensed SMGs. The cumulative number counts of the parent AS2COSMOS sample were presented in Simpson et al. (2020), and they are plotted in \autoref{fig:fig14}. Since our sample is limited to 12.4\,mJy at 870\,$\mu$m, the cumulative counts for the strongly lensed SMGs can be simply scaled from the parent counts of the corresponding flux bin (12.1\,mJy in this case) by dividing its value by our sample size. There is no source between 12.1\,mJy and 12.4\,mJy so no correction needs to be applied. To estimate the uncertainty, the Poisson error of 1 is adopted and propagated based on the uncertainty of the parent counts. The results are plotted in \autoref{fig:fig14} as the red point. As a check, the value is very close to one divided by the size of the footprint of the primary S2COSMOS catalog so one divided by 1.6 degree square. Overall we find again good agreement with the model prediction of Cai-Negrello.

\subsection{Connecting to the descendants}\label{sec:connection}
As shown in \autoref{sec:model}, we find that the emission lines of 15 out of the 17 ($\sim$90$\pm$30\%) primary SMGs with line detection appear to be better modeled by double Gaussian profiles. This high fraction of SMGs exhibiting double Gaussian line profiles is in contrast with previous studies of other SMGs samples, which have typically reported fractions under 50\% \citep{Greve:2005p6788,Bothwell:2013lp,Birkin:2021aa}. This could be in part due to our smaller sample size, or it could also be because of our sample SMGs being brighter so they may appear dynamically distinct from their fainter counterparts. Unless dispersion dominated, it is expected that the lines would appear double Gaussian in systems that exhibit velocity gradients, either due to orderly rotating disks or mergers. Spatially resolved dynamical studies are needed to make a more definite conclusion on their physical origins. 


\begin{figure}[h]
	\begin{center}
		\leavevmode
		\includegraphics[scale=0.8]{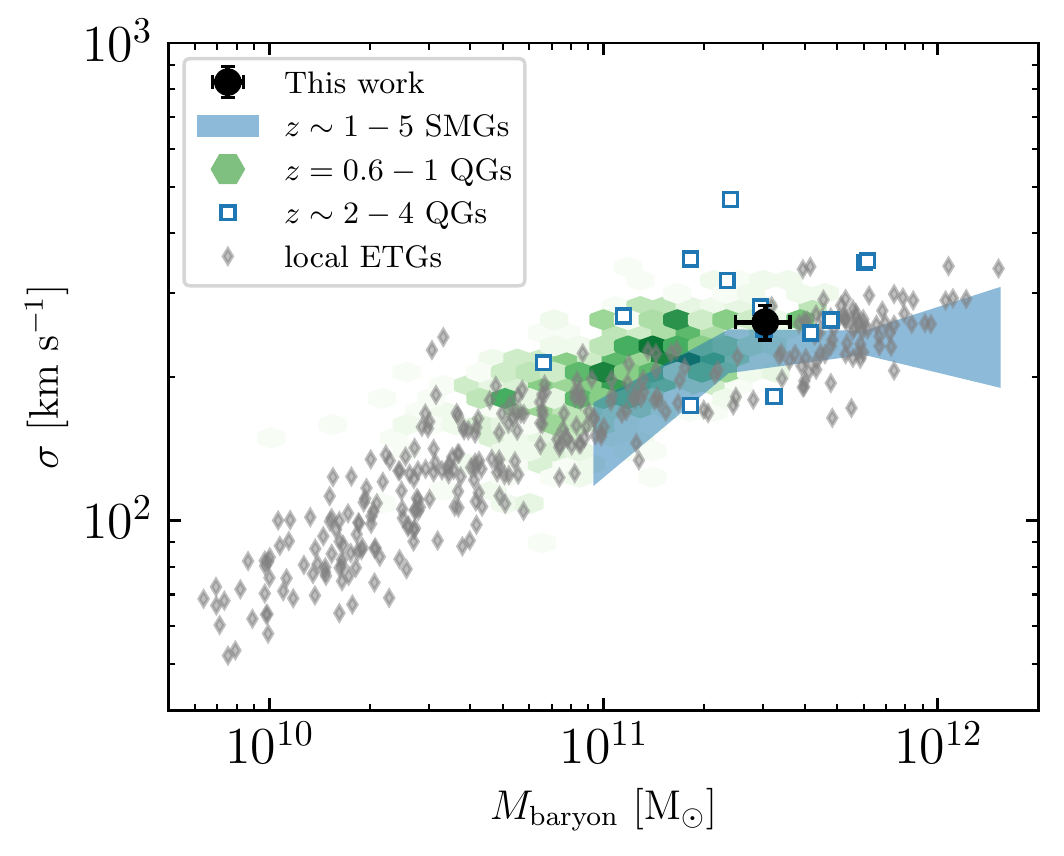}
		\caption{Correlations between baryon mass ($M_{\rm baryon}$) and velocity dispersion ($\sigma$). An overall median is shown in black circle for both parameters for our sample SMGs. A compiled sample of local ETGs including MASSIVE \citep{Davis:2019aa} and ATLAS$^{\rm 3D}$ \citep{Cappellari:2013aa} is plotted in gray, and recent measurements of $z=0.6-1.0$ and $z\sim2-4$ quiescent galaxies (QGs) are shown in green hexagons as 2D histogram \citep{van-der-Wel:2021aa,Wu:2021aa} and blue squares \citep{Stockmann:2020aa,Valentino:2020ab}, respectively. We also include results of fainter SMGs that also have spectroscopic redshift measurements \citep{Birkin:2021aa}. Our results are consistent with these plotted samples, corroborating the hypothesis that these populations are linked in evolution.
		}
		\label{fig:fig15}
	\end{center}
\end{figure}

On the other hand, we find that in our sample, for lines that are better described by double Gaussian, the median separation in velocity between the red and blue peaks is 430$\pm$40\,km\,s$^{-1}$, which is consistent with previous studies using their double-peaked SMGs \citep{Bothwell:2013lp,Birkin:2021aa}, as well as recent spatially resolved observations of CO/[CII] lines on a few SMGs \citep{Chen:2017aa,Calistro-Rivera:2018aa,Litke:2019aa}. This measurement of red and blue velocity separations allow us to crudely estimate dynamical masses provided some reasonable assumptions. Dynamical mass estimate can be expressed as $M_{\rm dyn} = C(v_c/{\rm sin}(i))^2\times R/G$, where $G$ is the gravitational constant, $v_c$ is the circular velocity, $R$ is the representative radius, $C$ is the correction factor depending on the mass distributions and the adopted representative radius, and $i$ is the inclination angle. This assumes a rotation dominated system, which is supported by recent high resolution and high sensitivity dynamical measurements of SMGs (e.g., \citealt{De-Breuck:2014aa,Lelli:2021aa,Rizzo:2021aa}). Assuming an exponential disk profile with a scale length $h$ and a half-light radius $r_e$, it has been shown that $R=3h=1.79r_e$\footnote{$r_e=1.68h$ for an exponential disk profile \citep{Chen:2015aa}} is a good approximation for spatially unresolved CO observations \citep{deBlok:2014aa}. Under this assumption of an exponential disk $C$ is about 2 \citep{Binney:2008aa}. We adopt $r_e=3$\,kpc based on recent spatially resolved CO measurements of SMGs \citep{Chen:2017aa,Calistro-Rivera:2018aa} and $v_c$ half of the median velocity separation that we measure. Since $v_c$ in our case is an average over a sample of sources with random inclination angles, we adopt an average $\langle{\rm sin}^2(i)\rangle=2/3$ \citep{Tacconi:2008p9334}. Since the dynamical mass estimated this way only includes mass within the assumed $r_e$, we multiply a factor of two to obtain the total dynamical mass. As a result, we estimate a total dynamical mass of $M_{\rm dyn} = 3.5\pm0.6\times10^{11} M_\odot$. Accounting for the typical dark matter fraction at high redshifts, which was recently reported as 12\% by \citet{Genzel:2020aa}, we deduce a total baryon mass of $M_{\rm baryon}=3.1\pm0.6\times10^{11} M_\odot$. {Note that the observational constraints on dark matter fraction are obtained from galaxies at $z\sim2$, slightly lower than our sample SMGs. However, recent Illustris-TNG hydrodynamical simulations have shown that the dark matter fractions are expected to be lower at higher redshifts \citep{Lovell:2018aa}. On the other hand, the simulations have predicted about a factor of two higher fractions than those obtained from observations. It is outside the scope of this paper to argue which is correct, although by adopting either a factor of two higher dark matter fraction or no dark matter our conclusions would not be significantly altered. Finally, in the following paper where we plan to present the properties of the ISM via CO/[CI] and dust SED measurement and modeling (Liao et al., in preparation), we will show that by including stellar, gas, and dust mass the averaged total baryon mass is consistent with the  dynamical estimates presented in this work.}

Following \citet{Birkin:2021aa}, The estimates of total baryon mass and the measurements of velocity dispersion via molecular lines allow us to assess the connection between our sample SMGs and their proposed descendants at lower redshifts, such as compact quiescent galaxies at similar redshifts or local massive early-type galaxies (ETGs; \citealt{Lilly:1999lr,Swinbank:2006aa,Alexander:2012aa,Toft:2014aa,Dudzeviciute:2020aa}). This is under the assumption that the total mass and dynamical energy are mostly preserved along the evolution. 

In \autoref{fig:fig15} we plot our sample as a combined median where $\sigma$ is the median of all the line moment measurements presented in \autoref{sec:model}, which is 260$\pm$21\,km\,s$^{-1}$. As a comparison measurements of local ETGs that are based on the MASSIVE \citep{Davis:2019aa} and ATLAS$^{\rm 3D}$ \citep{Cappellari:2013aa} surveys are shown, as well as the quiescent galaxies (D$_n$4000$>1.55$; \citealt{Kauffmann:2003aa,Wu:2018aa}) at $z=0.6-1.0$ based on the LEGA-C survey \citep{van-der-Wel:2021aa,Wu:2021aa} and some compiled $z\sim2-4$ spectroscopically confirmed quiescent galaxies \citep{Stockmann:2020aa,Valentino:2020ab}. The results from the similar analyses of mostly fainter SMGs presented by \citet{Birkin:2021aa} are also shown. For ETGs and high-redshift quiescent galaxies we adopt their stellar masses for the total baryon masses given stellar mass dominates baryon mass in these populations, whereas for fainter SMGs in \citet{Birkin:2021aa} stellar, gas, and dust masses are summed to deduce their baryon masses. Our sample of bright SMGs lies in the regions occupied by local ETGs and high-redshift quiescent galaxies, corroborating the proposed evolutionary link among these populations. 

In addition, given that our sample is well-defined in both flux density and area, we can compute number densities of our sample SMGs and compare the results with those based on recently confirmed quiescent galaxies at similar redshifts, especially those at $z>4$. The total number of our primary SMGs in the redshift bins of $2.1<z<3$, $3<z<4$, and $4<z<5.1$ are 6, 6, and 5, respectively. The lower bound of the lowest redshift bin is determined by the upper bound of the redshift gap (\autoref{sec:zdeter}), and the upper bound of the highest redshift bin is justified by the fact that given our wavelength coverage emission lines should have been detected from sources at $z=5.1-7.2$. The sky area covered by our initial SCUBA-2 survey is 1.6 deg$^2$ \citep{Simpson:2019aa} so the volumes covered in the three redshift bins are all approximately $2.0\times10^7$\,cMpc$^3$. Thus the observed number density of SMGs with $S_{870}>12.4$\,mJy is about $3\times10^{-7}$\,cMpc$^{-3}$ for all three redshift bins. However, assuming a typical lifetime of 100\,Myr for our sample SMGs (Liao et al. in preparation) the duty cycle corrected number densities are {$3.3^{+2.8}_{-1.8}\times10^{-6}$\,cMpc$^{-3}$, $1.9^{+1.6}_{-1.1}\times10^{-6}$\,cMpc$^{-3}$, and $1.0^{+0.9}_{-0.6}\times10^{-6}$\,cMpc$^{-3}$} for the three redshift bins, respectively. The uncertainties include poisson errors and a 40\% field-to-field variation \citep{Simpson:2020aa}. Our number density estimate at $z>4$ is similar to that presented by searches based on {\it Herschel} \citep{Ivison:2016aa} and ALMA 2\,mm survey \citep{Manning:2021aa}, although sample selections are different thus direct comparisons are not straightforward.

There has been an active discussion regarding the progenitors of recently confirmed $z\sim3-4$ quiescent galaxies, and extreme dusty star-forming galaxies such as SMGs at $z>4$ have been proposed to be ideal candidate given their nature (e.g., \citealt{Straatman:2014aa,Valentino:2020ab}). From the number density, our results at $z>4$ appears to be in agreement with those of $z\sim3-4$ quiescent galaxies estimated by \citet{Davidzon:2017aa} and \citet{Girelli:2019aa} ($\sim10^{-6}$\,cMpc$^{-3}$) but lower by a factor of a few to ten compared to other estimates ($\sim10^{-5}$\,cMpc$^{-3}$; \citealt{Straatman:2014aa,Schreiber:2018ac,Merlin:2019aa}). The different number densities estimated for quiescent galaxies could be due to selection effects such as different mass limits, as well as the methods to classify quiescent galaxies. However, given that we are selecting the brightest SMGs we expect our number density estimates to be lower limits, and fainter SMGs at $z>4$ are confirmed to exist (\autoref{sec:zvsobs}). We can conclude that our sample of SMGs have properties expected for the progenitors of some $z\sim3-4$ quiescent galaxies, especially the most massive ones with stellar mass of $>10^{11}$\,$M_\odot$.

\section{Summary} \label{sec:sum}
We report the first results of the AS2COSPEC survey, a millimeter spectroscopic survey using ALMA to target the brightest SMGs drawn from AS2COSMOS \citep{Simpson:2020aa}, a sample constructed through ALMA follow-up continuum observations of a flux-limited ($S_{\rm 850}>6.2$\,mJy) sample of submillimeter sources uncovered by the single-dish SCUBA-2 S2COSMOS survey \citep{Simpson:2019aa}. We targeted the brightest 18 primary SMGs with their flux densities between 12.4-19.3\,mJy at 870\,$\mu$m ($L_{\rm IR}\sim10^{13}$\,$L_\odot$; Liao et al. in preparation), along with five fainter companion SMGs that are located within the primary beam of our observations. The 18 primary SMGs represent a 100\% complete selection considering both the original SCUBA-2 and the follow-up ALMA observations. Our findings are the following.

\begin{enumerate}
	
	\item With about one hour of observations per target, we detect emission lines in CO or [CI] on 17 of the 18 primary SMGs. This high efficacy ($>90$\%) represents a major step forward in measuring spectroscopic redshifts of high-redshift luminous ($L_{\rm IR}>10^{12}$\,$L_\odot$) dusty galaxies that are otherwise difficult to obtain from traditional optical and near-infrared spectrographs of 8-10\,m class telescopes.
	
	\item Among the five companion SMGs we obtain significant line detection from one, AS2COS0008.2. Together with the previously reported line detection toward AS2COS0001.2, we find that the velocity differences of these two pairs are both within 500\,km\,s$^{-1}$, at the projected distances of 20-30\,kpc, suggesting that they are gravitationally bound systems therefore confirmed physically associated pairs. Our results suggest that for SMGs with $S_{870}>12.4$\,mJy, their brightest companion SMGs are physically associated with their corresponding primary SMGs in $\ge40$\% of the time, suggesting that mergers play a role in the triggering of star formation in these exceptional sources.
	
	\item The overall median redshift of the 18 primary SMGs is ${3.3\pm0.3}$, confirming that brighter SMGs are located at higher redshifts. {By carefully comparing with various model predictions we find that although the significance differs our measured redshift distribution is higher than the predictions of those models, where the only model made with a complete suite of physically motivated treatments has the most significant disagreement, at 3.8\,$\sigma$. We suspect that this may be because that the models lack halos with sufficiently massive gas reservoirs, which may be caused by either too efficient star formation or too much feedback at even earlier times.}
	
	\item By exploiting the rich ancillary data sets in the COSMOS field, we carefully model the foreground gravitational fields and find that only one of the 18 primary SMGs can be strongly lensed with a magnification $\mu>2$, suggesting a strongly lensed fraction of $<10$\%. Our results are consistent with previous studies focusing on {\it Herschel}-selected dusty galaxies, as well as the empirical model that interprets SMGs as proto-spheroidal galaxies. 

	\item With these high signal-to-noise spectra, we determine that about 90\% of the primary SMGs have their emission lines better described with double Gaussian profiles. The median separation of the two line peaks of the best-fit double Gaussian models is 430$\pm$40\,km\,s$^{-1}$, consistent with previous studies where redshifted and blueshifted components are spatially resolved. The average dynamical mass ($3.5\pm0.6\times10^{11}$\,$M_\odot$) estimated based on the velocity separation, together with the median line dispersion measured, puts our primary SMGs on similar mass-$\sigma$ correlations found on local ETGs and high-redshift quiescent galaxies, corroborating the proposed evolutionary link among these populations. 
	
	\item Lastly, given the well-defined nature of our sample we estimate that the number density of our primary SMGs at $z=4-5.1$ is $1^{+0.9}_{-0.6}\times10^{-6}$\,cMpc$^{-3}$, after accounting for the duty cycle for which we assume a typical lifetime of 100\,Myr. This suggests that our primary SMGs can be the progenitors of some $z\sim3-4$ quiescent galaxies, especially the most massive ones ($>10^{11}$\,$M_\odot$).
	
\end{enumerate}

We demonstrate the power of millimeter spectroscopic redshift surveys on a complete sample of bright SMGs in constraining theoretical models. In the forthcoming paper (Liao et al. in preparation) we will present the detailed analyses of the ISM properties of this effectively a well-defined sample of $z\sim3$ hyper luminous infrared galaxies (HyLIRGs; $L_{\rm IR}\gtrsim10^{13}$\,$L_\odot$). These efforts are in line with future developments of ALMA, ngVLA, and AtLAST \citep{Geach:2019aa,Klaassen:2020aa}, where submillimeter and millimeter spectroscopic surveys are expected to bring fundamental breakthroughs to our understandings of the high-redshift, dust-enshrouded galaxies.

\begin{acknowledgements}
{We thank the reviewer for a constructive report that improves the manuscript.} We would also like to thank Mathiew Bethermin, Caitlin Casey, Chris Hayward, Claudia Lagos, Chris Lovell, Mattia Negrello, Gergo Popping, Martin Sparre, and Jorge Zavala, for sharing their thoughts and catalogs about their models. We also thank Hsi-Wei Yen for the suggestions on the auto-masking approach, and Po-Feng Wu for sharing the stellar mass estimates of the LEGA-C sample. C.C.C. acknowledges support from the Ministry of Science and Technology of Taiwan (MOST 109-2112-M-001-016-MY3). IRS and AMS acknowledge support from STFC (ST/T000244/1). H. U. is supported by JSPS KAKENHI Grant No. 20H01953. AJB acknowledges funding from the “FirstGalaxies” Advanced Grant from the European Research Council (ERC) under the European Union’s Horizon 2020 research and innovation programme (Grant agreement No. 789056). The Cosmic Dawn Center (DAWN) is funded by the Danish National Research Foundation under Grant No. 140. This paper makes use of the following ALMA data: ADS/JAO.ALMA\#2019.1.01600.S. ALMA is a partnership of ESO (representing its member states), NSF (USA) and NINS (Japan), together with NRC (Canada), MOST and ASIAA (Taiwan), and KASI (Republic of Korea), in cooperation with the Republic of Chile. The Joint ALMA Observatory is operated by ESO, AUI/NRAO and NAOJ
\end{acknowledgements}
%

\vspace{5mm}
\facilities{ALMA}


\software{astropy \citep{Astropy-Collaboration:2013aa},
		{\sc casa} \citep{McMullin:2007aa}
          }



\bibliography{bib}{}
\bibliographystyle{aasjournal}

\appendix
\counterwithin{figure}{section}
\counterwithin{table}{section}

 \section{Spectral line model fitting}\label{appendix:linefitting}
To understand which model, single or double Gaussian, better describes the spectra, we employ the Bayesian information criterion (BIC) and a version of Akaike information criterion that is modified for the limited sample size (AICc), and they are defined as 
 
 \begin{equation*}
 	{\rm BIC} = kln(n)-2ln(\hat{L}) 
 \end{equation*}
 \begin{equation*}
 	{\rm AICc} = 2k+\frac{2k^2+2k}{n-k-1}-2ln(\hat{L})
 \end{equation*}	
 
 where $k$ is the number of parameters in the model and $n$ is the number of data points. $\hat{L}$ represents the maximized value of the likelihood function of the model, and in our case it is simply $\hat{L}={\rm exp}^{-\chi^2/2}$ where $\chi^2$ is the minimized $\chi^2$ value derived from the fitting. Note when $n\rightarrow\infty$ AICc becomes the original definition of AIC. In both cases, whichever model produces a lower value is expected to be closer to the truth. However, they each has its own limitations, in particular BIC typically imposes heavier penalty on additional parameters so it tends to be more conservative. Circumstances thus need to be considered to select the most optimal model. Our strategy is to make the selection based on the results of both criteria in combination with probability estimates from a set of Monte Carlo simulations. 

 The procedure is detailed in the following. All the measured spectra are first fit by both single and double Gaussian models\footnote{To ensure sensible results we impose a few loose boundaries; For each Gaussian component the peak has to be larger than 1\,$\sigma$, and the line widths need to be at least 100\,km\,s$^{-1}$. The separation of the two peaks for the double Gaussian models should be less than 1000\,km\,s$^{-1}$. As seen in \autoref{fig:fig6} these limits have negligible impact to most sources but in a few cases can help avoid fitting to noise peaks.}. The resulting reduced $\chi^2$ and the preferred model based on the two information criteria are presented in \autoref{table1}. For each line, the two best-fit models are then used to create two respective sets of mock spectra, where each set contains 500 mock spectra made by adding the corresponding best-fit model to the noise spectra drawn from 500 random positions. Each noise spectrum has the same frequency range as the measured spectrum, so this method has the advantage of preserving the noise structures, thus the weights for fitting, in the frequency domain (\autoref{fig:fig2}). Each mock spectrum is then fit by both single and double Gaussian models with exactly the same settings that are used to fit the measured spectra, and the selection of the preferred model is then performed and recorded. Finally we estimate the probability of a given line better described by a single or double Gaussian model by doing the following. First we select the mock spectra that have the preferred models matched to those of the measured spectrum, and among these selected mocks we compute the respective fractions of them made with the best-fit single or double Gaussian model. The two fractions computed this way are then the probability estimates of a given line being truly a single or double Gaussian line, and the results are show in \autoref{table1} in the columns of SGprob and DGprob. Note that these probabilities can change in different runs of simulations, but the variation is typically a few percents.
 
 For each line, our final selection of the preferred model then goes to the one that has a higher probability, except for AS2COS0001.1, AS2COS0023.1, and AS2COS0028.1, where our simulations yield inconclusive results either due to the comparable probabilities, or conflicting results from the two lines on the same target. \citet{Jimenez-Andrade:2020aa} reported CO(5-4) and [CII] detection on AS2COS0001.1, and with better signal-to-noise measurements they favored double-peak line profiles for both lines. We therefore adopt their results. For AS2COS0023.1, single Gaussian model is clearly preferred for CO(5-4). While double Gaussian model is preferred for CO(4-3), the bluer and fainter component is only marginally detected, with an velocity-integrated significance of $\sim2$\,$\sigma$ over the line FWHM (\autoref{fig:fig6}). We therefore adopt single Gaussian for AS2COS0023.1. Finally for AS2COS0028.1, although our simulations yield inconclusive results for both CO lines, the fact that the best-fit double Gaussian models have significantly different components suggest that the best fits are unavoidably subject to noise structures, especially for CO(4-3)\footnote{We have tried changing the boundaries for CO(4-3) but the broad component nevertheless appears to be in favor. It would result into significantly inconsistent second moment between the two lines, which is unlikely to be real.} (\autoref{fig:fig6}). Given their acceptable $\chi^2$ we therefore adopt single Gaussian models for AS2COS0028.1 for their consistency and simplicity. 
 
 As a check, we perform simultaneous fitting on SMGs that have two lines detected in our data. In this exercise, each source has its redshift fixed for the two lines, and for the double Gaussian models the velocity separations of the two constituent Gaussian are also fixed. We find the results are consistent with our decisions. The final selected models are given in the last column of \autoref{table1}, and they are further discussed in \autoref{sec:connection}.
 
  \begin{table*}
 	\caption{Best-fit model selection based on AICc and BIC}             
 	\label{table1}      
 	\centering
 	\hspace*{-2cm}\begin{tabular}{lrrllrrr}
 		\hline
 		ID & $\chi^2_{\text{SG}}$ & $\chi^2_{\text{DG}}$ & AICc & BIC & SGprob & DGprob & Selected Model\\
 		\hline
 		AS2COS0001.1 & 0.73 & 0.67 & DG & SG & 0.49 & 0.51 & DG\\
 		AS2COS0002.1 & 1.53 & 1.26 & DG & DG & 0.18 & 0.82 & DG\\
 		AS2COS0006.1 & 1.29 & 0.91 & DG & DG & 0.15 & 0.85 & DG\\
 		AS2COS0008.1 & 2.41 & 1.80 & DG & DG & 0.13 & 0.87 & DG\\
 		AS2COS0009.1 & 2.27 & 1.96 & DG & DG & 0.17 & 0.83 & DG\\
 		AS2COS0011.1-CI(1-0)& 1.20 & 1.04 & DG & DG & 0.16 & 0.84 & DG\\
 		AS2COS0011.1-CO(5-4) & 2.85 & 1.73 & DG & DG & 0.14 & 0.86 & DG\\
 		AS2COS0013.1 & 1.42 & 1.04 & DG & DG & 0.10 & 0.90 & DG\\
 		AS2COS0014.1 & 1.51 & 0.95 & DG & DG & 0.07 & 0.93 & DG\\
 		AS2COS0023.1-CO(4-3) & 1.33 & 1.17 & DG & DG & 0.27 & 0.73 & SG\\
 		AS2COS0023.1-CO(5-4) & 1.42 & 1.40 & SG & SG & 0.60 & 0.40 & SG\\
 		AS2COS0028.1-CO(3-2) & 0.96 & 0.84 & DG & SG & 0.51 & 0.49 & SG\\
 		AS2COS0028.1-CO(4-3) & 1.23 & 1.12 & DG & SG & 0.47 & 0.53 & SG\\
 		AS2COS0031.1-CO(4-3) & 4.21 & 1.81 & DG & DG & 0.19 & 0.81 & DG\\
 		AS2COS0031.1-CI(1-0) & 0.81 & 0.64 & DG & DG & 0.05 & 0.95 & DG\\
 		AS2COS0044.1 & 1.81 & 0.91 & DG & DG & 0.21 & 0.79 & DG\\
 		AS2COS0054.1 & 1.82 & 1.22 & DG & DG & 0.11 & 0.89 & DG\\
 		AS2COS0065.1 & 2.47 & 1.36 & DG & DG & 0.20 & 0.80 & DG\\
 		AS2COS0066.1 & 1.65 & 1.37 & DG & DG & 0.07 & 0.93 & DG\\
 		AS2COS0090.1 & 1.65 & 1.14 & DG & DG & 0.14 & 0.86 & DG\\
 		AS2COS0139.1 & 0.97 & 0.80 & DG & DG & 0.02 & 0.98 & DG\\
 		\hline
 	\end{tabular}
 	\hspace*{0cm}\begin{tabular}{l}
 		Note: SG and DG are abbreviations for single Gaussian and double Gaussian models. \\
 		The first and second columns after the ID column provide the reduced $\chi^2$ values of each fit, \\
 		and the last two columns show the preferred selection based on AICc and BIC, respectively.
 	\end{tabular}
 \end{table*}

 \section{Further comparisons between photometric redshifts and spectroscopic redshifts}\label{appendix:redshifts}
\begin{figure}[ht!]
	\begin{center}
		\leavevmode
		\includegraphics[scale=0.65]{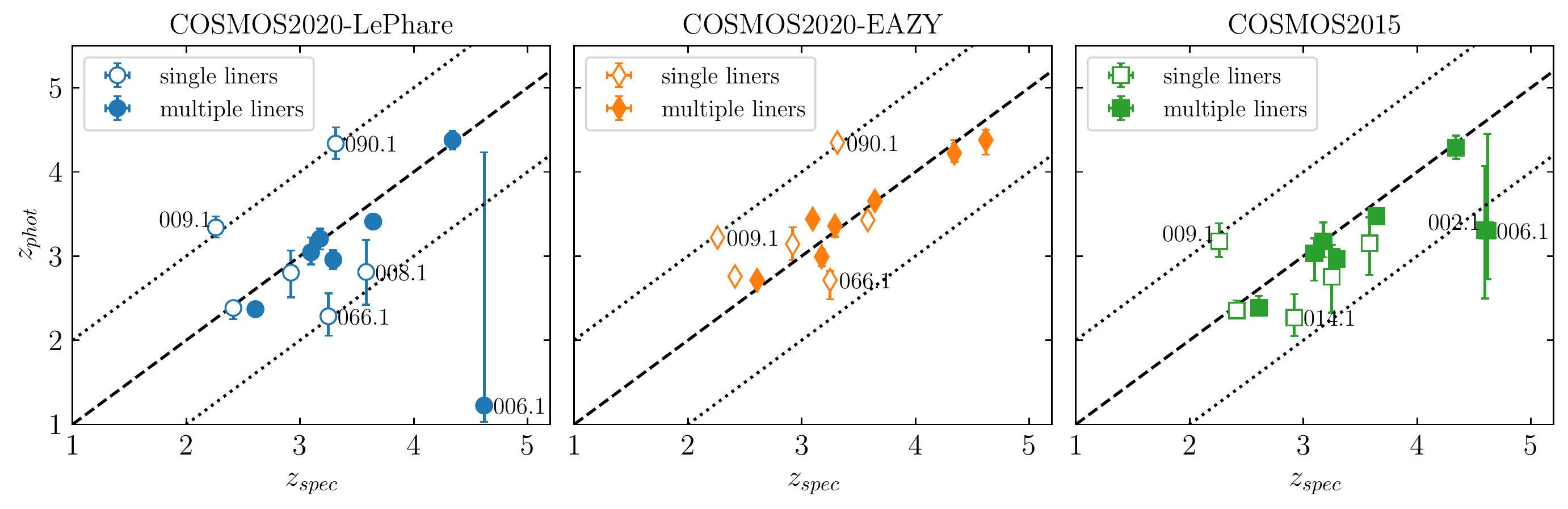}
		\caption{Similar to the top panel of \autoref{fig:fig7} where spectroscopic redshifts are taken from the adopted redshifts in \autoref{table2} and photometric redshifts are based on two COSMOS catalogs, COSMOS2015 \citep{Laigle:2016aa} and COSMOS2020 \citep{Weaver:2021aa}. Two versions of redshifts are provided in COSMOS2020 and they are both shown. 
		}
		\label{fig:A1}
	\end{center}
\end{figure}

 To gain more insights into the redshift assignment for the single line detected SMGs, we compare the adopted spectroscopic redshifts in \autoref{table2} and the photometric redshifts provided by the two well tested OIR-base catalogs in the COSMOS field, COSMOS2015 \citep{Laigle:2016aa} and COSMOS2020 \citep{Weaver:2021aa}. Focusing on the primary SMGs, matches to the two catalogs can be found on 14, where the corresponding matches are not exactly the same and most are found on SMGs with single line detection. The corresponding redshifts for the matched SMGs are plotted in \autoref{fig:A1} except for AS2COS0037.1 where no emission line is detected. Two versions of photometric redshifts are provided in the COSMOS2020 catalog and they are both shown. A few key features worth noting. First, the overall median in redshift difference is consistent with zero for all three sets of comparisons, suggesting that the redshift distribution discussed in \autoref{sec:redshift} is not affected by the photometric catalogs adopted for determining redshifts for single line SMGs. Second, compared to \autoref{fig:fig7}, OIR-base photometric redshifts tend to miss or underestimate redshifts for the highest redshift ($z>4$) SMGs. This is expected as they tend to be faint or undetected in OIR images. Finally, AS2COS0009.1 has significantly higher photometric redshifts ($\Delta z \gtrsim1$) in all three matches, suggesting that this source could be one of those that are expected to have incorrect redshift assignments in \autoref{table2}. Note AS2COS0090.1 can not be at the photometric redshifts deduced from the COSMOS2020 catalog, which is $z\sim4.3$, as two CO lines would have been detected, like AS2COS0023.1, given the frequency coverage of our data (\autoref{fig:fig5}).  
 
 
 
\end{document}